\documentclass[journal,10pt]{IEEEtran}

\usepackage{cite}
\usepackage{graphicx}
\usepackage{amsmath}
\usepackage{array}
\usepackage{mdwmath}
\usepackage{amssymb}
\usepackage{mdwtab}
\usepackage{stfloats}
\usepackage[tight,footnotesize]{subfigure}
\usepackage{amsmath,amsthm}
\usepackage{threeparttable}
\usepackage{color}
\usepackage{url}
\usepackage{algpseudocode}
\usepackage{algorithm}
\usepackage{multicol}
\usepackage{amssymb}
\usepackage{epstopdf}
\algrenewcommand{\algorithmicrequire}{\textbf{Input:}}
\algrenewcommand{\algorithmicensure}{\textbf{Output:}}

\newtheorem{theorem}{Theorem}
\newtheorem{remark}{Remark}
\newtheorem{lemma}{Lemma}

\begin{document}
\title{ Reconfigurable Intelligent Surface Aided Space Shift Keying With Imperfect CSI}
\author{ Xusheng Zhu, Wen Chen, \IEEEmembership{Senior Member, IEEE}, Qingqing Wu, \IEEEmembership{Senior Member, IEEE}, Zhendong Li,  \\Jun Li, \IEEEmembership{Senior Member, IEEE},  Shunqing Zhang, \IEEEmembership{Senior Member, IEEE}, and Ming Ding, \IEEEmembership{Senior Member, IEEE}

\thanks{
This paper has been presented in part at IEEE ICCC, in 2023 \cite{zhuper994}.
(\emph{Corresponding author: Wen Chen}).}
\thanks{X. Zhu, W. Chen, Q. Wu, and Z. Li are with the Department of Electronic Engineering, Shanghai Jiao Tong University, Shanghai 200240, China (e-mail: xushengzhu@sjtu.edu.cn; wenchen@sjtu.edu.cn; qingqingwu@sjtu.edu.cn; lizhendong@sjtu.edu.cn).}
\thanks{J. Li is with the School of Electronic and Optical Engineering, Nanjing University of Science Technology, Nanjing 210094, China (e-mail: jun.li@njust.edu.cn).}
\thanks{S. Zhang is with the Key Laboratory of Specialty Fiber Optics and Optical Access Networks, Shanghai University, Shanghai 200444, China (e-mail: shunqing@shu.edu.cn).}
\thanks{M. Ding is with Data61, CSIRO, Sydney, NSW 2015, Australia (e-mail:
ming.ding@data61.csiro.au).}
}

\maketitle

\begin{abstract}
In this paper, we investigate the performance of reconfigurable intelligent surface (RIS)-aided spatial shift keying (SSK) wireless communication systems with imperfect channel state information (CSI). Specifically, we study the average bit error probability (ABEP) of two RIS-SSK systems based on intelligent reflection and blind reflection modes.
For the intelligent RIS-SSK scheme, we first derive the conditional pairwise error probability of the composite channel through maximum likelihood (ML) detection.
Subsequently, we derive the probability density function of the combined channel.
Due to the intricacies of the composite channel formulation, an exact closed-form ABEP expression is unattainable through direct derivation. To this end, we resort to employing the Gaussian-Chebyshev quadrature method to estimate the results.
Additionally, we employ  Q-function approximation to derive the non-exact closed-form expression in the presence of channel estimation errors.
For the blind RIS-SSK scheme, we derive both closed-form ABEP expression and asymptotic ABEP expression with imperfect CSI by adopting the ML detector.
To offer deeper insights, we explore the impact of discrete reflection phase shifts on the performance of the RIS-SSK system.
Lastly, we extensively validate all the analytical derivations via Monte Carlo simulations.
\end{abstract}
\begin{IEEEkeywords}
Reconfigurable intelligent surface, space shift keying, imperfect channel state information, average bit error probability.
\end{IEEEkeywords}

\section{Introduction}
Reconfigurable intelligent surfaces (RISs) have recently attracted considerable concern due to their ability to make environments controllable \cite{wu2019towards}.
Particularly, RIS is an electromagnetic metasurface comprising small, low-cost, and almost passive scattering elements that can induce a predetermined phase shift in the incident wave \cite{li2022joi}.
Unlike conventional high energy consumption systems \cite{wang2017tc}, RIS can efficiently modify the scattering, reflection, and refraction of the environment cost-effectively, thereby improving the efficiency of wireless networks \cite{wu2020intelligent}.
In the existing literature, RIS-assisted communication systems are well-studied, especially concentrating on RIS with completely passive reflective elements that reflect and incident the signal in the desired direction only by applying low-power electronics \cite{di2020smart}.
In early RIS-aided transmission schemes \cite{saad2020vis}, the RIS is deployed to boost the system performance in two different scenarios: 1) RIS is mounted in the channel to maximize the received signal-to-noise ratio (SNR) by adjusting the reflected phase shifts; 2) RIS is considered as part of the transmitter and transfers information with the phase of its elements.
Furthermore, the authors of \cite{yang2020cov} provided a thorough theoretical analysis of the coverage of the RIS-assisted communication systems.

Index modulation (IM) has received considerable attention as a possible candidate for next-generation wireless systems \cite{li2023index}.
In particular, IM enhances spectral efficiency of the considered multiple-input multiple-output (MIMO) system by transmitting additional information bits through the fundamental components of the considered MIMO transmission scheme \cite{eabar2016index}.
This allows only a fraction of the energy-consuming resources to be activated at any given time, making IM a highly energy-efficient option. For this reason, IM is viewed as a promising technology for 6G systems \cite{li2023index,eabar2016index}.
Spatial modulation (SM) is a promising IM scheme for MIMO systems, where the IM technique is utilized for transmitting antenna indexes to transmit additional information bits \cite{mesleh2008spatial,zhu2021performance}.
To further improve the spectral efficiency, \cite{mesleh2015} proposed an SM-based quadrature SM scheme, which separates the information of the symbol domain into real and imaginary parts and hands them to two antennas for transmission, respectively.
To better emphasize spatial domain information, \cite{jegan2009space} investigated the space shift keying (SSK) scheme that neglects the symbol domain information of the SM.
Due to the large path loss in the millimeter wave (mmWave) band, it is difficult to guarantee the reliability of the received data by utilizing SM techniques for information transmission at each time slot. In this regard, \cite{zhu2023qua} proposed a new quadrature spatial scattering modulation scheme that exploits the hybrid beamforming instead of a single antenna in the SM.

\begin{table*}[t]
\centering
\caption{\small{Notations in this paper}}
\small
\begin{tabular}{|c|c|c|c|}
\hline Notations & Definitions & Notations & Definitions \\
\hline$N_{\mathrm{t}}$ & Number of transmit antenna & $N_{\mathrm{r}}$ & Number of receive antenna \\
\hline$n_{\mathrm{t}}$ & Transmit antenna index & $P_s$ & The average transmit power \\
\hline ${x}$ & The transmitted signal & ${y}$ & The received signal\\
\hline $\mathbb{C}^{m\times n}$ & The space of $m\times n$ matrics & $\rm{mod}(\cdot,\cdot)$ & The modulus operation \\
\hline $ {\rm diag}(\cdot)$ & Diagonal matrix operation  & {$\sim$} &``Distributed as"\\
\hline$\lfloor \cdot \rfloor$ & The floor operator & $|\cdot|$ & The absolute value operation \\
\hline $L$ & Number of RIS elements & $\lambda$ & Wavelength \\
\hline $\mathbf{\hat H}^{LoS}$ & line-of-sight (LoS) path & $\mathbf{\hat H}^{NLoS}$ &  Non-LoS (NLoS)paths \\
\hline $\mathcal{N}(\cdot,\cdot)$ & Real Gaussian distribution&$\mathcal{CN}(\cdot,\cdot)$ &  Complex Gaussian distribution\\
\hline $L_{\rm x}$ & Number of rows of RIS array & $L_{\rm y}$ & Number of columns of RIS array \\
\hline $\varphi_{\rm x}$& Azimuth angle of departure of RIS & $\varphi_{\rm y}$ & Elevation angle of departure of RIS \\
\hline$d$ & The half-wavelength spacing & $l$ & Reflective element index of RIS \\
\hline$(\cdot)^T$ & The transpose operator & $\otimes$ & Kronecker product \\
\hline$\Pr(\cdot)$ & Probability of the event occurring  & $P_b$ & CPEP \\
\hline$\kappa$ & The Rician factor & $\bar P_b$ & UPEP \\
\hline$\Re\{\cdot\}$ &Take the real part operation &$Var(\cdot)$ & The variance operator \\
\hline $\mathrm{E}(\cdot)$ & The expectation operation & $Q(\cdot)$ & The Q-function \\
\hline $f_X(\cdot)$ & PDF & $F_X(\cdot)$ & Cumulative distribution function (CDF) \\
\hline $\sin(\cdot)$ & Sine function & ${\rm sinc}(x)$ & $\sin(x)/x$ \\
\hline $\Phi(\cdot)$ & The Gaussian error function & ${\rm exp}(\cdot)$ & The Exponential function \\
\hline
\end{tabular}
\end{table*}

In light of the advantages possessed by RIS and IM, the RIS-assisted IM system has attracted extensive research interest from the academic community
\cite{easar2020reconf,ma2020large,can2020re,canbilen2022on,li2021space,singh2022ris,yuan2021receive,lin2022reconf,zhu2021ris,li2023int,li2022rec}.
Specifically, in \cite{easar2020reconf}, three RIS-based schemes are constructed based on SM/SSK techniques, i.e., applying SSK at the transmitter, RIS, and receiver sides, respectively, thus integrating IM into the RIS-assisted communication domain. It is worth mentioning that \cite{easar2020reconf} focuses on the analysis of the scheme for implementing SM/SSK at the receiver and uses average bit error probability (ABEP) as a metric.
In order to further enhance the spectral efficiency, \cite{ma2020large} applies SM technology to the transmitter and receiver sides, respectively.
In \cite{li2023int}, the authors discussed the application of RIS-SM in various scenarios and investigated the interaction between SM and RIS.
The authors of\cite{li2022rec} investigated the RIS-aided number modulation for symbiotic active/passive communications, where the RIS elements are divided into in-phase and quadrature subsets depending on their phase shift configurations.
Besides, the RIS-aided SSK scheme was presented in \cite{can2020re}, where the antenna is switched and selected at the transmitter side and the RIS is viewed as a passive relay.
In \cite{li2021space}, the RIS incorporates Alamouti space-time block coding, allowing the RIS to send its Alamouti-encoded data and reflect the incoming SSK signals toward the target.
To boost the spectral efficiency, \cite{canbilen2022on} studied the RIS-assisted full-duplex (FD) SSK system in the presence of the perfect self-interference (SI) cancellation.
However, in practical scenarios, it is challenging for SI to be perfectly eliminated cleanly. In view of this, \cite{singh2022ris} discussed the impact of the presence of residual SI on the RIS-assisted FD-SSK system.
Additionally, the authors of \cite{yuan2021receive} and \cite{lin2022reconf} investigated the RIS-aided receive quadrature reflecting modulation scheme, in which the entire RIS is logically partitioned into two halves to generate only in-phase and quadrature signals, each half forming a beam to a receiving antenna that carries the bit information utilizing the index of the antenna.
With the aim of studying RIS for mmWave information transmission, \cite{zhu2021ris,zhu2023risx,zhu2023ssmpe} evaluated the performance of RIS-assisted spatial scattering modulation (RIS-SSM) system in terms of reliability.

All the above-mentioned RIS-aided literature assumes that the channel state information (CSI) is completely well-known at the transceiver. Nevertheless, in reality, the estimated CSI is imperfect on account of estimation errors and limited radio resources of the RIS. To this end, it is significant to explore the impact of the imperfection of the estimated CSI (i.e., channel estimation error) on the RIS-aided system performance.
For imperfect CSI, \cite{zhou2020robust} investigated the worst-case robust beamforming design of RIS-assisted multi-user multiple-input single-output (MU-MISO) systems with the optimization objective of maximizing the transmit power.
In \cite{yang2022per}, the authors investigated the impact of an imperfect CSI on performance of the RIS-aided systems in terms of outage probability, average bit error rate, and average capacity.
Although work on RIS-assisted communication systems under imperfect CSI is common, there is no work on RIS-assisted IM schemes under imperfect CSI in the literature.
Against this background, we intend to elucidate this timely and interesting topic.
To the best of our knowledge, there is no analytical approach that has been adopted so far to investigate imperfect CSI for RIS-assisted SSK system error performance.
For clarity, the contribution of this paper are summarized as follows:
\begin{itemize}
\item
In this paper, we investigate RIS-assisted SSK downlink communication systems where the channel between the base station and RIS (BS-RIS) obeys Rayleigh fading, while the channel between RIS and user equipment (RIS-UE) obeys Rician fading. We consider perfect CSI estimation in the BS-RIS channel, but imperfect CSI estimation in the RIS-UE channel. We study both the intelligent RIS-SSK scheme and the blind RIS-SSK scheme and investigate estimation errors for both fixed and variable estimation errors.
\item
We use the maximum likelihood (ML) detection algorithm to analyze both schemes separately and derive conditional pairwise error probability (CPEP) expressions. Moreover, the complexity of the two detection schemes is also provided. By employing the central limit theorem (CLT), we derive the expectation and variance of the composite channel, and then obtain the probability density function (PDF)  for the intelligent and blind RIS phase-shift schemes, respectively.
\item
By combining the derived CPEP and PDF, we derive the closed-form expressions of the unconditional pairwise error probability (UPEP) in the presence of the imperfect CSI case. Furthermore, we derive asymptotic UPEP expression and ABEP expression. For the intelligent RIS-SSK scheme, the exact integration expression of ABEP is provided. Also,  we derive the exact estimation expression via the Gaussian-Chebyshev quadrature (GCQ) method and then derive the moment-generating function (MGF) expression. Additionally, we provide an analytical idea of RIS at uniform quantized phase shifts.
\item
We verify the ABEP expressions under the intelligent and blind RIS-SSK schemes using detailed Monte Carlo simulations, as well as the asymptotic expressions using analytical results, and the convergence and accuracy of the GCQ method. Our results demonstrate that the stronger the correlation between the estimated channel and the real received channel, or the higher the number of transmit pilots, the better the ABEP performance of the system. Furthermore, it is found that intelligent and blind RIS-SSK schemes represent the two extreme cases of RIS quantized phase shift.
\end{itemize}

The rest of the paper is organized as follows. Section II describes the intelligent and blind RIS-SSK system models for the RIS-SSK scheme under fixed and variable estimation error variances. In Section III, the analytical closed-form expressions for the intelligent and blind RIS-SSK schemes are derived. Besides, the discrete phase shift of the RIS scheme is also provided.
Section IV analyzes and discusses the numerical simulation and analytical results. Finally, Section V concludes the whole paper.
Notations in this paper are summarized in TABLE I.

\section{System Model}
In this section, we study the RIS-SSK system model under imperfect CSI, where the optimal reflection phase shift and blind reflection phase shift of the RIS are discussed.
In this regard, the SSK technique is employed at the Tx, where only one transmit antenna is activated at each time slot based on the input bits, while the other transmit antennas are in silent.
Note that the RIS tunes the phase shift by the information fed back from the BS side.

\begin{figure}[t]
  \centering
  \includegraphics[width=7cm]{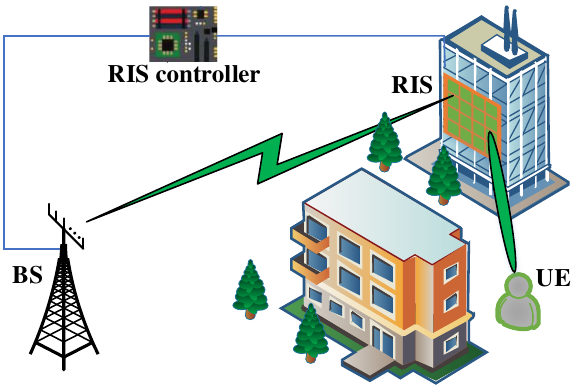}\\
  \caption{\small{System model of the RIS-SSK scheme.}}\label{systemmod}
  \label{sysim}
\end{figure}

It can be observed that Fig. \ref{sysim} consists of three components: the BS, the UE, and the RIS, where BS and UE are respectively equipped with $N_{\rm t}$ and one antennas \cite{can2020re}, and RIS is comprised of a two-dimensional uniform planar array (UPA) of $L = L_{\rm x} \times L_{\rm y}$ reflecting elements.
Based on \cite{abd2020ah}, RIS achieves a more substantial performance gain when positioned closer to the sides of the transceiver.
Accordingly, we deploy the RIS closer to the UE and farther away from the BS in Fig. \ref{sysim}.
Due to the single-antenna transmission mechanism of the SSK and the far distance from the BS to the RIS, we consider that the BS-RIS channel follows Rayleigh fading. On the other hand, the reflecting signal from the RIS can create a fine directional beam via large-scale elements, so we consider that the RIS-UE channel follows Rician fading.

Due to the existence of building, the direct link between the BS and UE is blocked in Fig. \ref{sysim}. To address this issue, we exploit RIS to assist information exchange between BE and UE.
It is worth noting that the RIS is attached to the exterior wall of the building, resulting in a relatively slow change channel between the BS and the RIS.
In views of this, the BS-RIS channel can be perfectly obtained via the channel estimation method provided in \cite{wei2021channel}.
Conversely, owing to the mobility of the UE, the channel between the RIS and UE can be changed rapidly.
Considering this, the RIS-UE channel is more challenging to acquire as the location of the UE and environmental factors change.

\subsection{Channel Model}
Due to the reflection enhancement of RIS, the RIS-UE channel $\mathbf{H}\in \mathbb{C}^{1\times L}$  can be modeled as
\begin{equation}
\mathbf{H} = \zeta \mathbf{\hat H} +\sqrt{1-\zeta^2}\Delta\mathbf{H},
\end{equation}
where $\zeta$ is the correlation coefficient between $\mathbf{H}$ and $\mathbf{\hat H}$, where $\mathbf{\hat H}$ is the information obtained by the channel estimation technique, and $\mathbf{H}$ stands for the practical channel obtained at the UE side. The corresponding estimation error is denoted by $\Delta\mathbf{H}$.
In particular, $\mathbf{\hat H}$ and $\Delta\mathbf{H}$ are mutually uncorrelated.
According to \cite{yang2022per}, the reflection estimation channel $\mathbf{H}$ can be formulated as
\begin{equation}\label{h1}
\mathbf{\hat H}=\sqrt{\frac{\kappa}{\kappa+1}}\mathbf{\hat H}^{LoS}+\sqrt{\frac{1}{\kappa+1}}\mathbf{\hat H}^{NLoS},
\end{equation}
where the deterministic LoS path matrix can be modeled as $\mathbf{\hat H}^{LoS}=\mathbf{\hat h}_{\rm x}(\varphi_{\rm x})\otimes \mathbf{\hat h}_{\rm y}(\varphi_{\rm y})$.
The row estimation vector $\mathbf{\hat h}_{\rm x}(\varphi_{\rm x})$ and the column estimation vector $\mathbf{\hat h}_{\rm y}(\varphi_{\rm y})$  respectively calculated as
\begin{subequations}
\begin{align}
&\mathbf{\hat h}_{\rm x}(\varphi_{\rm x}) = [1,e^{j\frac{2\pi d}{\lambda}\sin\varphi_{\rm x}},\ldots,e^{j\frac{2\pi d}{\lambda}(L_{\rm x}-1)\sin\varphi_{\rm x}}]^T,\\
&\mathbf{\hat h}_{\rm y}(\varphi_{\rm y}) = [1,e^{j\frac{2\pi d}{\lambda}\sin\varphi_{\rm y}},\ldots,e^{j\frac{2\pi d}{\lambda}(L_{\rm y}-1)\sin\varphi_{\rm y}}]^T,
\end{align}
\end{subequations}
where $l_{\rm x}={\rm mod}(l-1,L_{\rm x})$ and $ l_{\rm y}=\lfloor(l-1)/L_{\rm y}\rfloor$ represent the indices of the row and column of RIS, respectively.
It is worth noting that the fading channel between the $l$-th RIS reflecting shift and the receive antenna is denoted by $h_l=\beta_le^{-j\psi_l}$, where $\beta_l$ and ${\psi_l}$ stand for the amplitude and phase of the RIS-UE channel, respectively.
The expectation and variance of the magnitude of the path from the $l$-th reflecting element to the UE can be respectively expressed as \cite{san2007dig}
\begin{subequations}\label{beta}
\begin{align}
&E(\hat \beta_l)=\sqrt{\frac{\pi}{4\kappa+4}}e^{-\frac{\kappa}{2}}\left[(1+\kappa)
I_0\left(\frac{\kappa}{2}\right)+\kappa I_1\left(\frac{\kappa}{2}\right)\right],\\
&Var(\hat \beta_l) = 1-{ E}^2(\hat \beta_{l}),
\end{align}
\end{subequations}
where $I_a(\cdot)$ is the modified Bessel function of the first kind of order $a$.
For NLoS paths, each component of $\mathbf{\hat H}^{NLoS}$ suffers from $\mathcal{CN}(0,1)$.


On the other hand, the BS-RIS channel can be expressed as
$\mathbf{G}_{n_t}\sim \mathcal{CN}(0,\mathbf{I}_{L\times L})$, that is, the fading channel between the $n_t\in \{1,\cdots, N_t\}$ transmit antenna and the $l \in\{ 1,\cdots,L\}$ reflective element of the RIS
is represented by $g_{l,{n_t}}=\alpha_{l,n_t}e^{- j\theta_{l,n_t}}$,
where $\alpha_{l,n_t}$ and $\theta_{l,n_t}$ denote the amplitude and phase of the BS-RIS channel, respectively.
Accordingly, the mean and variance of the magnitude of the $l$-th reflecting element to $n_t$-th transmit antenna can be evaluated as \cite{san2007dig}
\begin{equation}\label{alpha}
\begin{aligned}
&E(\alpha_{n_t,l})=\frac{\sqrt{\pi}}{2}, \ \ \ Var(\alpha_{n_t,l}) = \frac{4-\pi}{4}.
\end{aligned}
\end{equation}

For the RIS, we set the amplitude of each reflection element of the RIS to one \cite{li2021space}.
Based on this, the reflection matrix of the RIS can be modeled as
\begin{equation}\label{phaseshif}
\boldsymbol{\Phi} = {\rm diag}(e^{j\phi_{1,n_t}},\cdots,e^{j\phi_{l,n_t}},\cdots,e^{j\phi_{L,n_t}}),
\end{equation}
where $e^{j\phi_{l,n_t}}$ denotes the phase shift that is related to the RIS controller connected to the $n_t$-th activated transmit antenna and the $l$-th reflecting element.

At the UE side, the received signal can be given as
\begin{equation}\label{y01}
y=\sqrt{P_s}\mathbf{ H}\boldsymbol{\Phi}\mathbf{G}_{n_t}x  + n_0,
\end{equation}
where $n_0\sim \mathcal{CN}(0,N_0)$ stands for the additive white Gaussian noise (AWGN).
Note that $x$ denotes the Gaussian data symbol, which is a random variable with zero mean and unit variance satisfying $E(|x|^2)=1$.
In this scheme, we aim to study the RIS-aided SSK technique, so the $x$ term can be neglected.
Hence, Eq. (\ref{y01}) can be reformulated as
\begin{equation}\label{y02}
y=\sqrt{P_s}\sum_{l=1}^Lh_le^{j\phi_{l,n_t}}g_{l,n_t}  + n_0,
\end{equation}
where $h_l = \zeta {\hat h} +\sqrt{1-\zeta^2}\Delta{h}$.
Further, Eq. (\ref{y02}) can be written as
\begin{equation}\label{y03}
\begin{aligned}
y=&\sqrt{P_s}\zeta\sum_{l=1}^L\hat h_le^{j\phi_{l,n_t}}g_{l,n_t} \\&+ \sqrt{P_s(1-\zeta^2)}\sum_{l=1}^L\Delta{h}e^{j\phi_{l,n_t}}g_{l,n_t} + n_0,
\end{aligned}
\end{equation}
where $\Delta h$ represents the error of channel estimation $\hat h$ of and obeys $\mathcal{CN}(0,\sigma_e^2)$ distribution. Particularly, $\sigma_e^2$ represents the variance of the estimation error, which depends on the estimation strategy and the number of pilot symbols employed \footnote{It is worth noting that $\sigma_e^2$ denotes the several factor on the CSI due to limited feedback and channel estimation.
Even in the high SNR region, the channel obtained at the UE is still inaccurate.}.
By adopting orthogonal pilot channel estimation sequences, the estimation error decreases linearly with the increase in the number of pilots.
According to \cite{basar2012per}, the correlation coefficient can be set as $\zeta = 1/\sqrt{1+\sigma_e^2}$.
It is worth mentioning that when $\sigma_e^2=0$, $\zeta = 1$ can be obtained, which indicates perfect channel estimation.

Note that this work considers two scenarios: one with a fixed $\sigma_e^2$ and another with a variable $\sigma_e^2$. In the fixed $\sigma_e^2$ scenario, the estimation error value remains constant across all SNR ranges. This is done to clarify the impact of imperfect channel information on the performance. In the variable $\sigma_e^2$ scenario, the estimation error value is related to the SNR through the formula $\sigma_e^2=N_0/(P_s N)$, where $N$ depends on the number of pilot symbols used and the chosen estimation method.

\subsection{Detection and Complexity Analysis}
\subsubsection{Detector for Intelligent RIS-SSK Scheme}
The RIS can adjust the phase shift to make $\phi_{l,n_t}=\theta_{l,n_t}+\psi_l$, thus maximizing the energy of the desired signal of the UE.
In this manner, the received signal can be demodulated by the ML detector, which can be given by
\begin{equation}\label{intdec}
[{\hat n_t}]=\arg\min\limits_{n_t\in\{1,\cdots,N_t\}}\left|y-\sqrt{P_s}\zeta\sum\nolimits_{l=1}^L\alpha_{l,{n_t}}\hat\beta_l\right|^2.
\end{equation}
\subsubsection{Detector for Blind RIS-SSK Scheme}
In this case, the RIS cannot tune the phase shifts, i.e., $\phi_{l,n_t}=0$, which is the worst case in terms of performance \cite{can2020re}.
In this respect, the ML detector is used to recover the original signal as follows:
\begin{equation}\label{bldec}
[{\hat n_t}]=\arg\min\limits_{n_t\in\{1,\cdots,N_t\}}\left|y-\sqrt{P_s}\zeta\sum\nolimits_{l=1}^L g_{l,n_t}\hat h_l\right|^2.
\end{equation}
\subsubsection{Complexity Analysis}
Note that every complex multiplication requires 4 real multiplications and 2 real additions. Computing the square of the absolute value of a complex number requires 2 real multiplications and 1 real addition. In  Eq. (\ref{intdec}), computing $\sum_{l=1}^{L}\alpha_{l,n_t}\beta_l$ requires $L$ real multiplications and $(L-1)$ real additions.
Computing $\sqrt{P_s}$ and $\zeta$ requires 2 real multiplications.
Subtracting $\sum_{l=1}^{L}\alpha_{l,n_t}\beta_l$ from $y$ requires 1 real addition. At this point, with $L+2$ real multiplications and $L$ real additions, to detect the transmitting antenna correctly, it is necessary to traverse and search through all the antennas on the transmission end. Therefore, the computational complexity of Eq. (\ref{intdec}) becomes $(L+4)N_t$ multiplications and $(L+1)N_t$ additions.

On the other hand, in Eq. (\ref{bldec}), computing $\sum_{l=1}^{L}h_{l,n_t} g_{l}$ requires performing $4L$ real multiplications and $4L-2$ real additions.
In particular, multiplication of $\sum_{l=1}^{L}h_{l,n_t} g_{l}$ and $\sqrt{P_s}$ $\zeta$ requires a total of 4 real multiplication operations.
Subtracting $\sum_{l=1}^{L}h_{l,n_t} g_{l}$ from $y$ requires performing 2 real additions.
In addition, when repeating this process for each transmitting antenna, the computational complexity of Eq. (\ref{bldec}) reaches $(4L+6)N_t$ multiplications and $(4L+1)N_t$ additions.

\section{Performance Analysis}
In this section, we derive the performance of the RIS-SSK scheme under imperfect CSI, where the RIS is used to connect the Rayleigh fading channel on the BS-RIS side and the Rician fading channel on the RIS-UE side. The CPEP and UPEP expressions for each scheme with the optimal ML detector are calculated. Furthermore, we obtain the ABEP expression of the RIS-SSK scheme with imperfect CSI.
\subsection{Error Probability for Intelligent RIS-SSK Scheme}
\subsubsection{CPEP}
It is assumed that the activated transmit antenna index is $n_t$ and the detected antenna index is $\hat n_t$. By exploiting the decision rules provided in Eq. (\ref{intdec}), the CPEP can be given as
\begin{equation}\label{xdfsg0}
\begin{aligned}
P_b =& \Pr\{n_t \to \hat{n}_t|\alpha_{l,{n_t}},\hat\beta_l\} \\
= &\Pr \{|y - \sqrt{P_s}\zeta\sum_{l=1}^L \alpha_{l,{n_t}}\hat\beta_l|^2 \\&>|y-\sqrt{P_s}\zeta\sum_{l=1}^L \alpha_{l,{\hat{n}_t}}\hat\beta_l e^{-j(\theta_{l,n_t}-\theta_{l,{\hat{n}_t}})}|^2\}\\
=&\Pr\{-2\Re\{y\sqrt{P_s}\zeta\sum_{l=1}^L \alpha_{l,{n_t}}\hat\beta_l\}+|\sqrt{P_s}\zeta\sum_{l=1}^L \alpha_{l,{n_t}}\hat\beta_l|^2 \\&>-2\Re\{y\sqrt{P_s}\zeta\sum_{l=1}^L \alpha_{l,{\hat{n}_t}}\hat\beta_l e^{-j(\theta_{l,n_t}-\theta_{l,{\hat{n}_t}})}\}\\&+|\sqrt{P_s}\zeta\sum_{l=1}^L \alpha_{l,{\hat{n}_t}}\hat\beta_l e^{-j(\theta_{l,n_t}-\theta_{l,{\hat{n}_t}})}|^2\}.
\end{aligned}
\end{equation}
To simplify the representation of Eq. (\ref{xdfsg0}), we can define
\begin{equation}\label{kap10}
\eta = \sum_{l=1}^L \alpha_{l,{n_t}}\hat\beta_l, \ \ \hat\eta=\sum_{l=1}^L \alpha_{l,{\hat{n}_t}}\hat\beta_l e^{-j(\theta_{l,n_t}-\theta_{l,{\hat{n}_t}})}.
\end{equation}
Substituting Eq. (\ref{kap10}) into Eq. (\ref{xdfsg0}), the CPEP can be updated to
\begin{equation}\label{sdfggsdg0}
\begin{aligned}
P_b
=&\Pr(-2\Re\{y\sqrt{P_s}\zeta\eta\}+|\sqrt{P_s}\zeta\eta|^2\\&>-2\Re\{y\sqrt{P_s}\zeta\hat \eta\}+|\sqrt{P_s}\zeta\hat \eta|^2)\\
=&\Pr(2\Re\{y\sqrt{P_s}\zeta(\hat\eta-\eta)\}\\&+|\sqrt{P_s}\zeta\eta|^2 -|\sqrt{P_s}\zeta \hat \eta|^2>0).
\end{aligned}
\end{equation}
Recall that Eq. (\ref{y02}), let us define $u = \sum_{l=1}^Lg_{l,n_t}e^{j\phi_{l,n_t}}\Delta h_{l}$.\footnote{For two independent random variables $X$ and $Y$, we can obtain the expectation and variance of term $XY$ as $E(XY)=E(X)E(Y)$ and $Var(XY)=Var(X)Var(Y)+Var(X)E^2(Y)+E^2(X)Var(Y)$, respectively.}
By adopting CLT, the $u$ obeys $\mathcal{CN}(0,\sigma_e^2L)$.
In this manner, Eq. (\ref{sdfggsdg0}) can be recast as
\begin{equation}\label{sdfggsdg1}
\begin{aligned}
P_b
=&\Pr(2\Re\{(\sqrt{P_s}\zeta\eta + \sqrt{P_s(1-\zeta^2)}u \\&+ n_0)\sqrt{P_s}\zeta(\hat\eta-\eta)\}+|\sqrt{P_s}\zeta\eta|^2  -|\sqrt{P_s}\zeta \hat \eta|^2>0)\\
=&\Pr(2\Re\{(\sqrt{P_s(1-\zeta^2)}u + n_0)\sqrt{P_s}\zeta(\hat\eta-\eta)\}\\&-|\sqrt{P_s}\zeta\eta|^2 -|\sqrt{P_s}\zeta \hat \eta|^2-2P_s\zeta^2\eta\hat\eta>0)\\
=&\Pr(2\Re\{(\sqrt{P_s(1-\zeta^2)}u + n_0)\sqrt{P_s}\zeta(\hat\eta-\eta)\}\\&-|\sqrt{P_s}\zeta\hat\eta-\sqrt{P_s}\zeta\eta|^2>0)\\
=&\Pr\left(D>0\right),
\end{aligned}
\end{equation}
where $D\sim\mathcal{N}(\mu_D,\sigma_D^2)$, where the expectation and variance of $D$ are represented as $\mu_D=-P_s\zeta^2|\hat\eta-\eta|^2$ and $\sigma_D^2={2(N_0+P_s(1-\zeta^2)\sigma_e^2L)}$, respectively.
In this respect, the Eq. (\ref{sdfggsdg1}) can be evaluated as
\begin{equation}\label{enu01}
\begin{aligned}
P_b
=&\Pr(-\mu_D/\sigma_D)=Q\left(\sqrt{\frac{P_s\zeta^2|\hat\eta-\eta|^2}{2(N_0+P_s(1-\zeta^2)\sigma_e^2L)}}\right).
\end{aligned}
\end{equation}
\subsubsection{UPEP}
By employing Eq. (\ref{kap10}), $\eta -\hat \eta$ can be written as
\begin{equation}\label{enu02}
\eta -\hat \eta = \sum\nolimits_{l=1}^L \hat\beta_l \left(\alpha_{l,{{n}_t}}- \alpha_{l,{\hat{n}_t}} e^{-j\omega}\right).
\end{equation}
where $\omega=\theta_{l,n_t}-\theta_{l,{\hat{n}_t}}$.
Since $\theta_{l,n_t}$ and $\theta_{l,{\hat{n}_t}}$ both independently and uniformly distributed in $(0,2\pi)$, then the PDF of $\omega$ can be given as follows:
\begin{equation}\label{eqphix}
f_{\omega}(x)=\left\{
\begin{aligned}
&\frac{1}{2\pi}(1+\frac{x}{2\pi}),\ \ \  x \in [-2\pi,0),\\
&\frac{1}{2\pi}(1-\frac{x}{2\pi}), \ \ \ x \in [0,2\pi).\\
\end{aligned}
\right.
\end{equation}
In this manner, the $\alpha_{l,{\hat{n}_t}} e^{-j\omega}$ in Eq. (\ref{enu02}) can be calculated as
\begin{equation}
\alpha_{l,{\hat{n}_t}} e^{-j\omega}=\alpha_{l,{\hat{n}_t}}\cos\omega-j\alpha_{l,{\hat{n}_t}}\sin\omega.
\end{equation}
Since the symmetry of cosine and sine function, we have
\begin{subequations}
\begin{align}
&E[\alpha_{l,{\hat{n}_t}} e^{-j\omega}] = 0, \\ &Var[(\alpha_{l,{\hat{n}_t}} e^{-j\omega})_\Re] = 1/2, \\ &Var[(\alpha_{l,{\hat{n}_t}} e^{-j\omega})_\Im] = 1/2.
\end{align}
\end{subequations}
It is known that the real and imaginary parts are two independent parts of each other, thus the variance of $\alpha_{l,{\hat{n}_t}} e^{-j\omega}$ is $Var[(\alpha_{l,{\hat{n}_t}} e^{-j\omega})] = 1$.

After some simple algebraic operations, we can derive the mean and variance of $\alpha_{l,{{n}_t}}- \alpha_{l,{\hat{n}_t}} e^{-j\omega}$ in Eq. (\ref{enu02}) as follows:
\begin{subequations}
\begin{align}
&E(\alpha_{l,{{n}_t}}- \alpha_{l,{\hat{n}_t}} e^{-j\omega}) = {\sqrt{\pi}}/{2},\\
&Var(\alpha_{l,{{n}_t}}- \alpha_{l,{\hat{n}_t}} e^{-j\omega}) = ({8-\pi})/{4}.
\end{align}
\end{subequations}
Further, the mean and variance of $\beta_l \left(\alpha_{l,{{n}_t}}- \alpha_{l,{\hat{n}_t}} e^{-j\omega}\right)$ in Eq. (\ref{enu02}) can be respectively expressed as
\begin{subequations}
\begin{align}
&E[\hat\beta_l (\alpha_{l,{{n}_t}}- \alpha_{l,{\hat{n}_t}} e^{-j\omega})] = {\sqrt{\pi}}E(\hat\beta_l)/{2},\\
&Var[\hat\beta_l (\alpha_{l,{{n}_t}}- \alpha_{l,{\hat{n}_t}} e^{-j\omega})] =2-{\pi}E^2(\hat\beta_l)/{4}.
\end{align}
\end{subequations}
Since the reflecting elements are independent of each other, it is difficult to directly obtain the accurate PDF concerning the sum of the composite channel. To address this issue, we use the CLT to approximate the PDF as a real Gaussian distribution. Consequently, the corresponding mean and variance can be respectively expressed as
\begin{equation}\label{musigma}
\mu = {\sqrt{\pi}LE(\hat\beta_l)}/{2},\ \
\sigma^2 = {L[8-\pi E^2(\hat\beta_l)]}/{4}. 
\end{equation}
\begin{figure}[t]
\centering
\includegraphics[width=8cm]{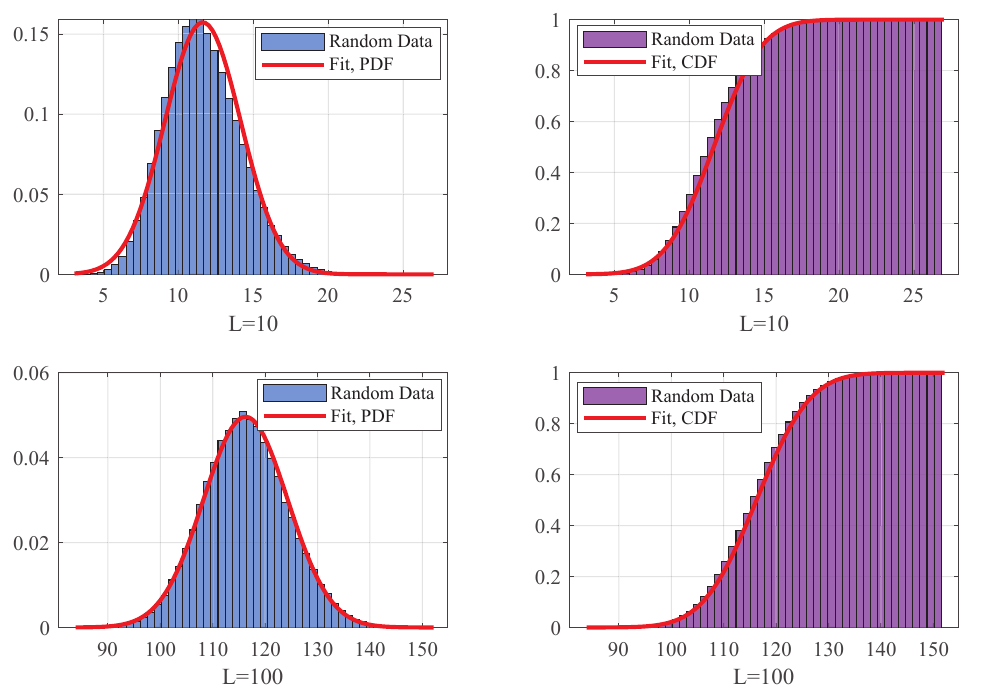}
\caption{\small{Fitting the PDF and CDF distributions of the data by using CLT.}}
\vspace{-10pt}
\label{sys}
\end{figure}
To understand more intuitively the gap between the PDF and CDF fitted using CLT and the reality PDF and CDF, we adopt $1\times10^5$ number of simulations in Fig. \ref{sys} for the BS-RIS side Rayleigh fading channel and RIS-UE side Rician fading channel, where the Rician factor is 3 dB.
It is worth mentioning that we plotted the number of RIS reflecting elements for $L=10$ and $L=100$ cases, respectively.
In Fig. \ref{sys}, it can be observed that there are some gaps between the PDF obtained by simulation and the CDF and the curve obtained by CLT fitting at $L=10$, while the two coincide nearly perfectly at $L=100$.

Based on Eqs. (\ref{enu01}) and (\ref{musigma}), the UPEP of the proposed scheme can be calculated as
\begin{equation}\label{q1}
\begin{aligned}
    \bar P_b &= \int_0^\infty Q\left(\sqrt{\frac{\rho\zeta^2x}{2(1+\rho(1-\zeta^2)\sigma_e^2L)}}\right)f(x)dx,
\end{aligned}
\end{equation}
where $x = |\eta-\hat\eta|^2$, $\rho={P_s}/{N_0}$ stands for SNR, and $f(x)$ denotes the PDF of $x$ variable.
Substituting
$
Q(x) = \frac{1}{\pi}\int_0^{\frac{\pi}{2}}\exp\left(-\frac{x^2}{2\sin^2\o}\right)d\o
$
into Eq. (\ref{q1}), the UPEP can be updated as
\begin{equation}
\begin{aligned}
    \bar P_b&=\frac{1}{\pi}\int_0^\infty \int_0^{\frac{\pi}{2}}\\&\times\exp\left(-\frac{\rho\zeta^2x}{4\sin^2\o(1+\rho(1-\zeta^2)\sigma_e^2L)}\right) f(x)d\o dx.
\end{aligned}
\end{equation}
By swapping the order of integration of $\o$ and $x$, we have
\begin{equation}\label{pep1}
\begin{aligned}
    \bar P_b &=\frac{1}{\pi} \int_0^{\frac{\pi}{2}} \int_0^\infty \\&\times \exp\left(-\frac{\rho\zeta^2x}{4\sin^2\o(1+\rho(1-\zeta^2)\sigma_e^2L)}\right) f(x) dx d\o,
\end{aligned}
\end{equation}
where $x$ obeys the non-central chi-square distribution with one degree. To address this issue, we resort the following lemma to get the PDF of $x$.
\begin{lemma}
The PDF of a non-central chi-square distribution with one degree of freedom can be expressed as
\begin{equation}\label{pdf}
    f_X(x) = \frac{\exp\left(-\frac{\mu^2}{2\sigma^2}\right)}{2\sqrt{2\pi\sigma^2 x}}\left[\exp\left(\frac{\mu\sqrt{x}}{\sigma^2}\right)+\exp\left(-\frac{\mu\sqrt{x}}{\sigma^2}\right)\right].
\end{equation}
Proof: Please refer to Appendix A.
$\hfill\blacksquare$
\end{lemma}
To obtain the closed-form expression of UPEP, we substitute Eq. (\ref{pdf}) into Eq. (\ref{pep1}). At this point, the (\ref{pep1}) can be represented as
\begin{equation}\label{pdfpdfw}
\begin{aligned}
  \bar  P_b
    =&\frac{\exp\left(-\frac{\mu^2}{2\sigma^2}\right)}{2\pi\sqrt{2\pi\sigma^2 }} \int_0^{\frac{\pi}{2}}\int_0^\infty \frac{1}{\sqrt{x}}\\&\times{\exp\left(-\frac{2({1+\rho(1-\zeta^2)\sigma_e^2L})\sin^2\o+\rho\zeta^2\sigma^2}{4({1+\rho(1-\zeta^2)\sigma_e^2L})\sigma^2\sin^2\o}x\right)}\\ &\times\left[\exp\left(\frac{\mu\sqrt{x}}{\sigma^2}\right)+\exp\left(-\frac{\mu\sqrt{x}}{\sigma^2}\right)\right] dx d\o,
\end{aligned}
\end{equation}
Let us make $t = \sqrt{x}$, the Eq. (\ref{pdfpdfw}) can be rewritten as
\begin{equation}\label{pep2}
\begin{aligned}
  \bar  P_b
    =&\frac{\exp\left(-\frac{\mu^2}{2\sigma^2}\right)}{\pi\sqrt{2\pi\sigma^2 }} \int_0^{\frac{\pi}{2}}\int_0^\infty \\&\times {\exp\left(-\frac{2({1+\rho(1-\zeta^2)\sigma_e^2L})\sin^2\o+\rho\zeta^2\sigma^2}{4({1+\rho(1-\zeta^2)\sigma_e^2L})\sigma^2\sin^2\o}t^2\right)}
    \\&\times\left[\exp\left(\frac{\mu t}{\sigma^2}\right)+\exp\left(-\frac{\mu t}{\sigma^2}\right)\right] dt d\o.
\end{aligned}
\end{equation}
On this basis, we further rewrite Eq. (\ref{pep2}) as
\begin{equation}\label{pep3}
\begin{aligned}
  &\bar  P_b
    =\frac{\exp\left(-\frac{\mu^2}{2\sigma^2}\right)}{\pi\sqrt{2\pi\sigma^2 }} \int_0^{\frac{\pi}{2}}\int_0^\infty  \\&\times {\exp\left(-\frac{2({1+\rho(1-\zeta^2)\sigma_e^2L})\sin^2\o+\rho\zeta^2\sigma^2}{4({1+\rho(1-\zeta^2)\sigma_e^2L})\sigma^2\sin^2\o}t^2+\frac{\mu }{\sigma^2}t\right)}
    \\&+\exp\left(-\frac{2({1+\rho(1-\zeta^2)\sigma_e^2L})\sin^2\o+\rho\zeta^2\sigma^2}{4({1+\rho(1-\zeta^2)\sigma_e^2L})\sigma^2\sin^2\o}t^2-\frac{\mu }{\sigma^2}t\right) dt d\o.
\end{aligned}
\end{equation}
To address the inner integral of Eq. (\ref{pep3}),
we resort to \cite{xx2007tab}
\begin{equation}\label{tab207_1}
\int_0^\infty\exp\left(-\frac{t^2}{4\delta}-\gamma t\right)dt=\sqrt{\pi \delta}\exp(\delta\gamma^2)[1-\Phi(\gamma\sqrt{\delta})].
\end{equation}
By applying Eq. (\ref{tab207_1}), we can update Eq. (\ref{pep3}) as
\begin{equation}\label{pep4}
\begin{aligned}
    \bar P_b&=\frac{\exp\left(-\frac{\mu^2}{2\sigma^2}\right)}{\pi\sqrt{2\sigma^2 }} \int_0^{\frac{\pi}{2}}\sqrt{\frac{{({1+\rho(1-\zeta^2)\sigma_e^2L})\sigma^2\sin^2\o}}{{2({1+\rho(1-\zeta^2)\sigma_e^2L})\sin^2\o+\rho\zeta^2\sigma^2}}}
    \\&\times\exp\left({\frac{{\mu^2({1+\rho(1-\zeta^2)\sigma_e^2L})\sigma^2\sin^2\o}}{{2\sigma^4({1+\rho(1-\zeta^2)\sigma_e^2L})\sin^2\o+\rho\zeta^2\sigma^6}}}\right)
    \\&\times\left[2-\Phi\left(-\sqrt{{\frac{{\mu^2({1+\rho(1-\zeta^2)\sigma_e^2L})\sigma^2\sin^2\o}}{{2\sigma^4({1+\rho(1-\zeta^2)\sigma_e^2L})\sin^2\o+\rho\zeta^2\sigma^6}}}}\right)\right.\\&\left.-\Phi\left(\sqrt{{\frac{{\mu^2({1+\rho(1-\zeta^2)\sigma_e^2L})\sigma^2\sin^2\o}}{{2\sigma^4({1+\rho(1-\zeta^2)\sigma_e^2L})\sin^2\o+\rho\zeta^2\sigma^6}}}}\right)\right] d\o.
\end{aligned}
\end{equation}
Herein, we adopt the following theorem to tackle the Eq. (\ref{pep4}).
\begin{theorem}
According to Eq. (\ref{pep4}), it is observed that the included variables of the $\Phi(\cdot)$ function are complex and for this reason, we deal with them as
\begin{equation}
\begin{aligned}
&\Phi\left(-\sqrt{{\frac{{\mu^2({1+\rho(1-\zeta^2)\sigma_e^2L})\sigma^2\sin^2\o}}{{2\sigma^4({1+\rho(1-\zeta^2)\sigma_e^2L})\sin^2\o+\rho\zeta^2\sigma^6}}}}\right)
\\&+\Phi\left(\sqrt{{\frac{{\mu^2({1+\rho(1-\zeta^2)\sigma_e^2L})\sigma^2\sin^2\o}}{{2\sigma^4({1+\rho(1-\zeta^2)\sigma_e^2L})\sin^2\o+\rho\zeta^2\sigma^6}}}}\right) = 0.
\end{aligned}
\end{equation}
{Proof:}Please refer to Appendix B.
$\hfill\blacksquare$
\end{theorem}
Based on the {\bf Theorem 1}, we can obtain the Eq. (\ref{pep4}) as
\begin{equation}\label{pep6}
\begin{aligned}
  &\bar  P_b
   =\frac{\sqrt{2}}{\pi{\sigma }} \int_0^{\frac{\pi}{2}}\sqrt{{\frac{{({1+\rho(1-\zeta^2)\sigma_e^2L})\sigma^2\sin^2\o}}{{2({1+\rho(1-\zeta^2)\sigma_e^2L})\sin^2\o+\rho\zeta^2\sigma^2}}}}\\&\times\exp\left({\frac{{\mu^2({1+\rho(1-\zeta^2)\sigma_e^2L})\sigma^2\sin^2\o}}{{2\sigma^4({1+\rho(1-\zeta^2)\sigma_e^2L})\sin^2\o+\rho\zeta^2\sigma^6}}}-\frac{\mu^2}{2\sigma^2}\right) d\o.
\end{aligned}
\end{equation}
By observing Eq. (\ref{pep6}), we can see that the integral variable $\o$ exists in the coefficient and exponential terms of the exp function, respectively. Also, the coefficient and exponential terms are very complicated with respect to $\o$, it is difficult to solve directly using the conventional method.
To address this issue, we adopt the GCQ method and the Q-function estimation method to obtain the exact estimated solution and the closed-form expression of Eq. (\ref{pep6}), respectively.

{\bf a) GCQ Method:}
To facilitate the implementation of the GCQ method, we set $\o=\frac{\pi}{4}\vartheta+\frac{\pi}{4}$. Thus, the Eq. (\ref{pep6}) can be given as (\ref{pepx6}), at the top of next page.
\begin{figure*}[t]
\begin{equation}\label{pepx6}
\begin{aligned}
  \bar  P_b
  & =\frac{\sqrt{2}}{4{\sigma }} \int_{-1}^1\sqrt{{\frac{{({1+\rho(1-\zeta^2)\sigma_e^2L})\sigma^2\sin^2\left(\frac{\pi}{4}\vartheta+\frac{\pi}{4}\right)}}{{2({1+\rho(1-\zeta^2)\sigma_e^2L})\sin^2\left(\frac{\pi}{4}\vartheta+\frac{\pi}{4}\right)+\rho\zeta^2\sigma^2}}}}\\&\times\exp\left({\frac{{\mu^2({1+\rho(1-\zeta^2)\sigma_e^2L})\sigma^2\sin^2\left(\frac{\pi}{4}\vartheta+\frac{\pi}{4}\right)}}{{2\sigma^4({1+\rho(1-\zeta^2)\sigma_e^2L})\sin^2\left(\frac{\pi}{4}\vartheta+\frac{\pi}{4}\right)+\rho\zeta^2\sigma^6}}}-\frac{\mu^2}{2\sigma^2}\right) d\vartheta.
\end{aligned}
\end{equation}
\hrulefill
\end{figure*}
After that, the integral form of Eq. (\ref{pepx6}) is rewritten as a summation form. Accordingly, the UPEP can be further characterized as Eq. (\ref{eqGCQ1}), shown at the top of the next page,
\begin{figure*}[t]
\begin{equation}\label{eqGCQ1}
\begin{aligned}
  \bar  P_b = & \frac{\sqrt{2}\pi}{4\sigma K}\sum_{k=1}^K\sqrt{{\frac{{(1-\vartheta_k^2)({1+\rho(1-\zeta^2)\sigma_e^2L})\sigma^2\sin^2\left(\frac{\pi}{4}\vartheta_k+\frac{\pi}{4}\right)}}{{2({1+\rho(1-\zeta^2)\sigma_e^2L})\sin^2\left(\frac{\pi}{4}\vartheta_k+\frac{\pi}{4}\right)+\rho\zeta^2\sigma^2}}}}\\&\times\exp\left({\frac{{\mu^2({1+\rho(1-\zeta^2)\sigma_e^2L})\sigma^2\sin^2\left(\frac{\pi}{4}\vartheta_k+\frac{\pi}{4}\right)}}{{2\sigma^4({1+\rho(1-\zeta^2)\sigma_e^2L})\sin^2\left(\frac{\pi}{4}\vartheta_k+\frac{\pi}{4}\right)+\rho\zeta^2\sigma^6}}}-\frac{\mu^2}{2\sigma^2}\right) +R_K,
\end{aligned}
\end{equation}
\hrulefill
\end{figure*}
where $K$ denotes the complexity-accuracy trade-off factor, $\vartheta_k = \cos\frac{2k-1}{2K}\pi$, and $R_K$ denotes the error term which can be ignored for high values of $K$.

{\bf b) Q-function Approximating Method:}
With the help of the approximation of the Q-function in \cite{zhu2023qua}, it can be indicated as
\begin{equation}\label{qapr}
Q(x) \approx \frac{1}{12}\exp\left(-\frac{x^2}{2}\right)+\frac{1}{4}\exp\left(-\frac{2x^2}{3}\right).
\end{equation}
Substituting Eq. (\ref{qapr}) into Eq. (\ref{q1}), the UPEP can be evaluated as
\begin{equation}\label{qapr1}
\begin{aligned}
\bar P_b
&\approx \frac{1}{12}\int_0^\infty \exp\left(-{\frac{\rho\zeta^2x}{4(1+\rho(1-\zeta^2)\sigma_e^2L)}}\right)f(x)dx \\&+\frac{1}{4}\int_0^\infty\exp\left(-{\frac{\rho\zeta^2x}{3(1+\rho(1-\zeta^2)\sigma_e^2L)}}\right)f(x)dx \\
= &\frac{1}{12}\int_0^\infty \exp\left(-{\frac{\rho\zeta^2x}{4(1+\rho(1-\zeta^2)\sigma_e^2L)}}\right)\\&\times\frac{\exp\left(-\frac{x+\mu^2}{2\sigma^2}\right)}{2\sqrt{2\pi\sigma^2 x}}\left[\exp\left(\sqrt{\frac{\mu^2x}{\sigma^4}}\right)+\exp\left(-\sqrt{\frac{\mu^2x}{\sigma^4}}\right)\right]dx \\&+\frac{1}{4}\int_0^\infty\exp\left(-{\frac{\rho\zeta^2x}{3(1+\rho(1-\zeta^2)\sigma_e^2L)}}\right)\\&\times\frac{\exp\left(-\frac{x+\mu^2}{2\sigma^2}\right)}{2\sqrt{2\pi\sigma^2 x}}\left[\exp\left(\sqrt{\frac{\mu^2x}{\sigma^4}}\right)+\exp\left(-\sqrt{\frac{\mu^2x}{\sigma^4}}\right)\right]dx.
\end{aligned}
\end{equation}
At this step, we provide the following lemma to address Eq. (\ref{qapr1}).
\begin{lemma}
It can be observed that Eq. (\ref{qapr1}) contains the form of MGF, therefore, Eq. (\ref{qapr1}) can be further obtained by matching to MGF as
\begin{equation}\label{closepb1}
\begin{aligned}
\bar P_b
\approx&\frac{1}{12}\sqrt{\frac{2(1+\rho(1-\zeta^2)\sigma_e^2L)}{2(1+\rho(1-\zeta^2)\sigma_e^2L)+\rho\zeta^2\sigma^2}}\\&\times\exp\left(\frac{-\rho\zeta^2\mu^2}{4(1+\rho(1-\zeta^2)\sigma_e^2L)+2\rho\zeta^2\sigma^2}\right)\\
&+\frac{1}{4}\sqrt{\frac{3(1+\rho(1-\zeta^2)\sigma_e^2L)}{3(1+\rho(1-\zeta^2)\sigma_e^2L)+2\rho\zeta^2\sigma^2}}\\&\times\exp\left(\frac{-\rho\zeta^2\mu^2}{3(1+\rho(1-\zeta^2)\sigma_e^2L)+2\rho\zeta^2\sigma^2}\right).
\end{aligned}
\end{equation}
Proof: Please refer to Appendix C.
$\hfill\blacksquare$
\end{lemma}

\subsubsection{Asymptotic UPEP}
By utilizing Chernov bound: $Q(x) \leq \frac{1}{2}\exp\left(-\frac{x^2}{2}\right)$, we have
\begin{equation}\label{closeupbound}
\begin{aligned}
\bar P_b
&\leq\frac{1}{6}\sqrt{\frac{2(1+\rho(1-\zeta^2)\sigma_e^2L)}{2(1+\rho(1-\zeta^2)\sigma_e^2L)+\rho\zeta^2\sigma^2}}\\&\times\exp\left(\frac{-\rho\zeta^2\mu^2}{4(1+\rho(1-\zeta^2)\sigma_e^2L)+2\rho\zeta^2\sigma^2}\right).
\end{aligned}
\end{equation}
To facilitate the analysis of the asymptotic performance of Eq. (\ref{closeupbound}), let us define
\begin{equation}\label{eq5}
\tau = \frac{\rho\zeta^2}{2(1+\rho(1-\zeta^2)\sigma_e^2L)}.
\end{equation}
Recall that $\zeta = 1/\sqrt{1+\sigma_e^2}$, Eq. (\ref{eq5}) can be re-expressed as
\begin{equation}
\tau =\frac{\frac{\rho}{1+\sigma_e^2}}{2\left(1+\frac{\rho\sigma_e^4L}{1+\sigma_e^2}\right)}=\frac{\rho}{2(1+\sigma_e^2+\rho\sigma_e^4L)}.
\end{equation}
Taking the limit operation for $\tau$, we have
$
\lim\limits_{\rho\to \infty}\tau = \frac{1}{2\sigma_e^4L}.
$
After some manipulations, Eq. (\ref{closeupbound}) can be updated as
\begin{equation}\label{closeupbound1}
\bar P_b
\leq\sqrt{\frac{2\sigma_e^4}{8\sigma_e^4+8-\pi E^2(\hat\beta_l)}}\exp\left({\frac{-\pi LE^2(\hat\beta_l)}{{16\sigma_e^4}+{16-2\pi E^2(\hat\beta_l)}}}\right).
\end{equation}
\begin{remark}
In the fixed $\sigma_e^2$ scenario, we observe that the Eq. (\ref{closeupbound1}) is the constant value.
\end{remark}
\begin{remark}
In the variable $\sigma_e^2$ scenario, the Eq. (\ref{closeupbound1}) is a variable value.
Recall that $\sigma_e^2=1/(\rho N)$, the Eq. (\ref{closeupbound1}) can be given as
\begin{equation}\label{closeupbound2}
\bar P_b
\leq\sqrt{\frac{\frac{2}{\rho^2N^2}}{\frac{8}{\rho^2N^2}+8-\pi E^2(\hat\beta_l)}}\exp\left({\frac{-\pi LE^2(\hat\beta_l)}{{\frac{16}{\rho^2N^2}}+{16-2\pi E^2(\hat\beta_l)}}}\right).
\end{equation}
When the SNR takes the limit, Eq. (\ref{closeupbound2}) can be characterized as
\begin{equation}\label{closeupbound3}
\bar P_b
\leq\frac{1}{2}\exp\left({\frac{-\pi LE^2(\hat\beta_l)}{{16-2\pi E^2(\hat\beta_l)}}}\right).
\end{equation}
In this case, Eq. (\ref{closeupbound3}) goes through two deflation transformation operations and the resulting expression is also a fixed value.
\end{remark}

\subsection{Error Probability for Blind RIS-SSK Scheme}
\subsubsection{CPEP}
In this subsection, we investigate the UPEP expression of the blind RIS-SSK with the imperfect CSI scenario.
By combining Eqs. (\ref{y03}) and (\ref{bldec}), the CPEP can be given as
\begin{equation}\label{xdfsg}
\begin{aligned}
P_b =& \Pr\{n_t \to \hat{n}_t|g_{l,n_t}\hat h_l\}\\
=& \Pr \{|y - \sqrt{P_s}\zeta\sum_{l=1}^L g_{l,n_t}\hat h_l|^2 >|y-\sqrt{P_s}\zeta\sum_{l=1}^L g_{l,{\hat n}_t}\hat h_l |^2\}\\
=&\Pr\{-2\Re\{y\sqrt{P_s}\zeta\sum_{l=1}^L g_{l,{n_t}}\hat h_l\}\\&+|\sqrt{P_s}\zeta\sum_{l=1}^L g_{l,{n_t}}\hat h_l|^2
>-2\Re\{y\!\sqrt{P_s}\zeta\sum_{l=1}^L g_{l,{\hat{n}_t}}\hat h_l \}\\&+|\sqrt{P_s}\zeta\sum_{l=1}^L g_{l,{\hat{n}_t}}\hat h_l |^2\}.
\end{aligned}
\end{equation}
Without loss of generality, let us define
\begin{equation}\label{kap1}
\eta = \sum_{l=1}^L g_{l,{n_t}}\hat h_l, \ \ \hat\eta=\sum_{l=1}^L g_{l,{\hat{n}_t}}\hat h_l,\ \ u=\sum_{l=1}^Lg_{l,n_t}\Delta h_{l}.
\end{equation}
Substituting Eq. (\ref{kap1}) into Eq. (\ref{xdfsg}), the $P_b$ can be recast as
\begin{equation}\label{sdfggsdg}
\begin{aligned}
P_b
=&\Pr(-2\Re\{y\sqrt{P_s}\zeta\eta\}+|\sqrt{P_s}\zeta\eta|^2\\&>-2\Re\{y\sqrt{P_s}\zeta\hat \eta\}+|\sqrt{P_s}\zeta\hat \eta|^2)\\
=&\Pr(2\Re\{y\sqrt{P_s}\zeta(\hat\eta-\eta)\}+|\sqrt{P_s}\zeta\eta|^2 -|\sqrt{P_s}\zeta \hat \eta|^2>0)\\
=&\Pr(2\Re\{(\sqrt{P_s}\zeta\eta + \sqrt{P_s(1-\zeta^2)}u + n_0)\sqrt{P_s}\zeta(\hat\eta-\eta)\}\\&+|\sqrt{P_s}\zeta\eta|^2 -|\sqrt{P_s}\zeta \hat \eta|^2>0)\\
=&\Pr(2\Re\{(\sqrt{P_s(1-\zeta^2)}u + n_0)\sqrt{P_s}\zeta(\hat\eta-\eta)\}\\&-{P_s}\zeta^2|\eta- \hat \eta|^2>0)\\
=&\Pr(D>0),
\end{aligned}
\end{equation}
where $D = 2\Re\{(\sqrt{P_s(1-\zeta^2)}u + n_0)\sqrt{P_s}\zeta(\hat\eta-\eta)\}-{P_s}\zeta^2|\eta- \hat \eta|^2$ following $\mathcal{N}(\mu_D,\sigma_D^2)$.
Therefore, the expectation and variance of the variable $D$ can be respectively represented as
 $\mu_D=-{P_s}\zeta^2|\eta- \hat \eta|^2$,
$\sigma_D^2=2 [P_s(1-\zeta^2)\sigma_e^2L+N_0]{P_s}\zeta^2|\eta- \hat \eta|^2$.
Based on this, the CPEP in Eq. (\ref{kap1}) can be reproduced as
\begin{equation}\label{sdfxxx1}
P_b = Q\left(\sqrt{{\mu_D^2}/{\sigma_D^2}}\right)=Q\left(\sqrt{\frac{\rho\zeta^2|\eta- \hat \eta|^2}{2 (\rho(1-\zeta^2)\sigma_e^2L+1)}}\right).
\end{equation}
It is obvious that $|\eta-\bar\eta|^2$ plays a dominant role in Eq. (\ref{sdfxxx1}). Consequently, we should address its corresponding distribution.
According to Eq. (\ref{kap1}), the $\eta$ consists of the summation of $L$ monomials, where each monomial is obtained by multiplying $g_{l,{n_t}}$ and $\hat h_l$.
For $\hat h_l$, we have
\begin{equation}
\hat h_l=\sqrt{\frac{\kappa}{\kappa+1}}e^{j\phi_{l,n_t}}+\sqrt{\frac{1}{\kappa+1}}\hat h_l^{NLoS},
\end{equation}
where
$\hat h_l^{NLoS} \sim \mathcal{CN}(0,1)$. As a result, we obtain
$\sqrt{\frac{1}{\kappa+1}}\hat h_l^{NLoS} \sim \mathcal{CN}\left(0,{\frac{1}{\kappa+1}}\right)$. Further, we can get $\hat h_l\sim \mathcal{CN}\left(\sqrt{\frac{\kappa}{\kappa+1}},{\frac{1}{\kappa+1}}\right)$.
Recall that $g_{l,n_t} \sim \mathcal{CN}(0,1)$, in accordance with the principle of multiplication of two independent variables, we can obtain that $g_{l,{n_t}}\hat h_l$ obeys $\mathcal{CN}(0,1)$.
Since each phase shift unit of RIS works independently, $\mathcal{CN}(0,L)$ can be obtained by using CLT.

\begin{remark}
With the above analysis, we find an interesting phenomenon that after the blind reflection of the Rayleigh fading channel and Rician fading channel by RIS, the composite channel is equivalent to a Rayleigh fading channel, independent of the Rician factor in the Rician fading channel.
\end{remark}

By using CLT, we have $\eta \sim \mathcal{CN}(0,L)$.
Similarly, we can obtain $\hat \eta \sim \mathcal{CN}(0,L)$.
Taking the rule of addition by employing independent random variables, we have
\begin{equation}\label{ricdem}
\begin{aligned}
&\eta-\hat \eta \sim \mathcal{CN}(0,2L).
\end{aligned}
\end{equation}
Let us define $x = |\eta-\bar\eta|^2$, the central chi-square PDF with two degrees of freedom can be represented as
\begin{equation}\label{blindpdf}
f(x)=\frac{1}{2L}\exp\left(-\frac{x}{2L}\right).
\end{equation}

\subsubsection{UPEP}
With the combination of Eqs.  (\ref{sdfxxx1}) and (\ref{blindpdf}), the UPEP can be computed by
\begin{equation}\label{Pb1b}
\begin{aligned}
\bar P_b = \int_0^\infty f(x)Q\left(\sqrt{\frac{\rho\zeta^2x}{2 \left(\rho(1-\zeta^2)\sigma_e^2L+1\right)}}\right)dx.
\end{aligned}
\end{equation}
By applying $Q(x) = \frac{1}{\pi}\int_0^{\frac{\pi}{2}}\exp\left(-\frac{x^2}{2\sin^2\o}\right)d\o$, the UPEP in Eq. (\ref{Pb1b}) can be evaluated as
\begin{equation}\label{Pb2b}
\begin{aligned}
\bar P_b &= \frac{1}{2\pi L}\int_0^\infty\int_0^{\frac{\pi}{2}} \exp\left(-\frac{x}{2L}\right)\\&\times\exp\left(-{\frac{\rho\zeta^2x}{4 \left(\rho(1-\zeta^2)\sigma_e^2L+1\right)\sin^2\o}}\right)d\o dx.
\end{aligned}
\end{equation}
After exchanging the order of integration of variables $\o$ and $x$, the Eq. (\ref{Pb2b}) can be reformulated as
\begin{equation}\label{Pb2b1}
\begin{aligned}
\bar P_b
&= \frac{1}{2\pi L}\int_0^{\frac{\pi}{2}} \int_0^\infty \exp\left(-\frac{x}{2L}\right)\\&\times\exp\left(-{\frac{\rho\zeta^2x}{4 \left(\rho(1-\zeta^2)\sigma_e^2L+1\right)\sin^2\o}}\right) dx d\o.
\end{aligned}
\end{equation}
Addressing the inner integral, we obtain
\begin{equation}\label{Pb2b2}
\begin{aligned}
\bar P_b
&= \frac{1}{\pi }\int_0^{\frac{\pi}{2}}  {\frac{2 \left(\rho(1-\zeta^2)\sigma_e^2L+1\right)\sin^2\o}{\rho L\zeta^2+{2 \left(\rho(1-\zeta^2)\sigma_e^2L+1\right)\sin^2\o}}}  d\o.
\end{aligned}
\end{equation}
To facilitate subsequent analysis, we rewrite Eq. (\ref{Pb2b2}) as
\begin{equation}\label{Pb2b3}
\begin{aligned}
\bar P_b
&= \frac{1}{\pi }\int_0^{\frac{\pi}{2}}  {\frac{\sin^2\o}{\frac{\rho L\zeta^2}{2 \left(\rho(1-\zeta^2)\sigma_e^2L+1\right)}+{\sin^2\o}}}  d\o.
\end{aligned}
\end{equation}
After some mathematical operations, the UPEP in Eq. (\ref{Pb2b3}) can be further expressed as
\begin{equation}\label{Pb2b4}
\begin{aligned}
\bar P_b
&=\frac{1}{2}\left(1-\sqrt{\frac{{\rho L\zeta^2}}{{\rho L\zeta^2}+{2 \left(\rho(1-\zeta^2)\sigma_e^2L+1\right)}}}\right).
\end{aligned}
\end{equation}
\subsubsection{Asymptotic UPEP}
In the high SNR region, the asymptotic UPEP can be described as
$
\bar P_{\rm asy}=\lim\limits_{\rho \to\infty}\bar P_b.
$
When the obtained power $\zeta$ of the desired signal is higher than the estimation error power $\sqrt{1-\zeta}$, i.e., $\zeta\gg1-\zeta$.
Here, we have
\begin{equation}\label{asyblind2}
\begin{aligned}
\bar P_{\rm asy}
&= \lim\limits_{\rho \to\infty}\frac{1}{\pi }\int_0^{\frac{\pi}{2}}  {\frac{\sin^2\o}{\frac{\rho L\zeta^2}{2 \left(\rho(1-\zeta^2)\sigma_e^2L+1\right)}+{\sin^2\o}}}  d\o\\&
= \lim\limits_{\rho \to\infty}\frac{1}{\pi }\int_0^{\frac{\pi}{2}}  {\frac{\sin^2\o}{\frac{\rho L\zeta^2}{2 \left(\rho(1-\zeta^2)\sigma_e^2L+1\right)}}}  d\o.
\end{aligned}
\end{equation}
After some mathematical calculations, Eq. (\ref{asyblind2}) can be organized as
\begin{equation}\label{asyblind3}
\begin{aligned}
\bar P_{\rm asy}
&= \lim\limits_{\rho \to\infty}{\frac{2 \left(\rho(1-\zeta^2)\sigma_e^2L+1\right)}{\rho L\zeta^2}}\times\frac{1}{\pi }\int_0^{\frac{\pi}{2}}  {\sin^2\o}  d\o.
\end{aligned}
\end{equation}
Refer to the Wallis Formula provided in \cite{xx2007tab}, Eq. (\ref{asyblind3}) can be reproduced as
\begin{equation}\label{asybli}
\begin{aligned}
\bar P_{\rm asy}
&=\lim\limits_{\rho \to\infty}{\frac{\rho(1-\zeta^2)\sigma_e^2L+1}{2\rho L\zeta^2}} = {\frac{(1-\zeta^2)\sigma_e^2}{2\zeta^2}}=\frac{\sigma_e^4}{2}.
\end{aligned}
\end{equation}

\subsection{Discrete Phase Shift of RIS}
In practical systems, it is challenging to achieve continuous phase shifts and adds considerable complexity to the system. To address this problem, discrete phase shifts are adopted. Concretely, each reflecting element can only obtain a value from a finite set of discrete phase shift values $2^Q$, which can be characterized as
\begin{equation}
\phi_{1,n_t}\in[0:2^Q-1]\frac{2\pi}{2^Q}-\pi+\frac{2\pi}{2^{Q+1}},
\end{equation}
where $Q$ denotes uniformly quantized phase shift levels. In particular, the discrete phase shift of each reflecting element is its continuous phase shift value to the nearest $\phi_{1,n_t}$ in the quantization of the point. It is worth mentioning that the quantization errors are uniformly distributed in the $\left[-\frac{\pi}{2^Q},\frac{\pi}{2^Q}\right]$ interval.
Accordingly, the reflecting phase shift in Eq. (\ref{phaseshif}) is set as
\begin{equation}
\begin{aligned}
\phi_{l,n_t} &= \frac{2^Q}{2\pi}\int_{-\frac{\pi}{2^Q}}^{\frac{\pi}{2^Q}}e^{jx}dx= \frac{2^Q}{2\pi j}\int_{-\frac{\pi}{2^Q}}^{\frac{\pi}{2^Q}}de^{jx}\\&
= \frac{2^Q}{2\pi j}\left(e^{j\frac{\pi}{2^Q}}-e^{j\frac{\pi}{2^Q}}\right)= \frac{2^Q}{\pi }\sin\frac{\pi}{2^Q}= {\rm sinc}\left(\frac{\pi}{2^Q}\right).
\end{aligned}
\end{equation}

\begin{remark}
In this subsection, we find that the intelligent RIS-SSK and blind RIS-SSK schemes are two special cases with respect to the quantization scheme. When the quantization bits are zeros, the quantization scheme evolves into the blind RIS-SSK case, while when the quantization bits are taken to infinity, the quantization scheme becomes the intelligent RIS-SSK case.
\end{remark}
\subsection{ABEP Expression}
It is worth noting that ABEP is equal to UPEP when $N_t$ is two, while ABEP is the joint upper bound of the scheme when $N_t$ is greater than two. Consequently, the ABEP of the RIS-SSK scheme can be characterized as
\begin{equation}
ABEP \leq \frac{1}{\log_2N_t}\sum_{\hat n_t=1}^{N_t}\sum_{n_t=1}^{N_t} \bar P_i N(\hat n_t\to n_t),
\end{equation}
where $i \in \{b, {\rm asy}\}$ and  $N(\hat n_t\to n_t)$ indicates the number of error bits between the true transmit antenna index $n_t$ and the decoded judgment obtained antenna index $\hat n_t$.

\section{Simulation and Analytical Results}
In this section, we investigate the error performance of the proposed scheme under imperfect CSI via Monte Carlo simulation. The simulation involves generating a random data sequence and transmitting it to the receiver via RIS reflection after modulation.
Unless otherwise specified, the simulation results of the each ABEP value corresponding to the SNR is generated via  $1\times10^6$ times, and $N_t$ and $N_r$ are respectively set to 2 and 1. Note that the impact of any large-scale path loss is neglected as it is already implicit in the received SNR.

\begin{figure}[t]
\begin{minipage}[t]{0.49\linewidth}
\centering
\includegraphics[width=4.5 cm]{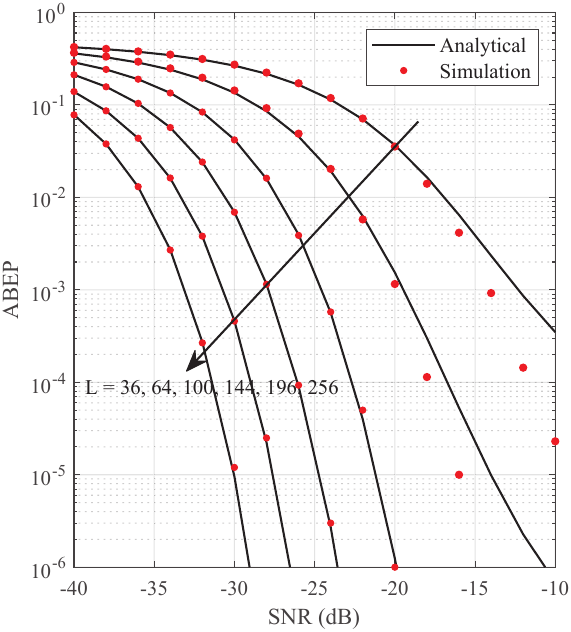}
\caption{\small{Verification of the conditions for the application of CLT under the intelligent RIS-SSK scheme.}}
\label{verfsa}
\end{minipage}%
\hfill
\begin{minipage}[t]{0.49\linewidth}
\centering
\includegraphics[width=4.5 cm]{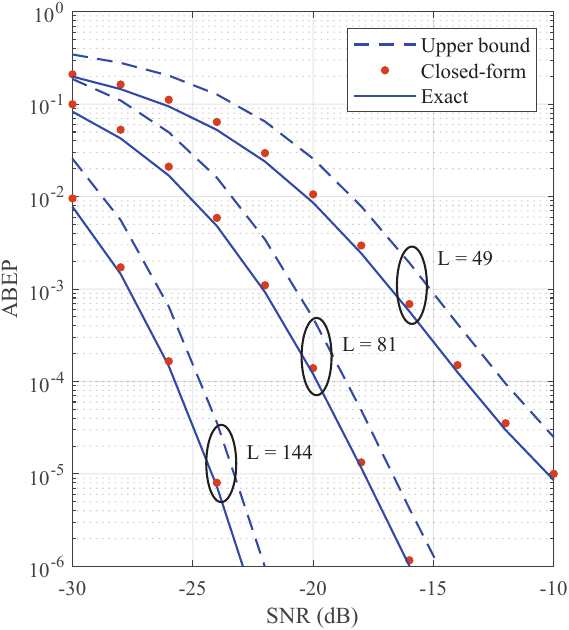}
\caption{\small{Validation of analytical derivation results under intelligent RIS-SSK scheme.}}
\label{verupbound}
\end{minipage}
\end{figure}
In Fig. \ref{verfsa}, we plot the ABEP performance of the RIS-SSK scheme with perfect CSI, where both sides of the RIS are set to follow the Rayleigh fading channels, i.e., the error estimation parameters $\sigma_e^2$ and the Rician factor are set to be zero.
It can be observed from Fig. \ref{verfsa} that when $L$ is relatively large, the agreement of the simulation results with the analytical curve is subsequently improved.
The reason of this phenomenon is that the analytical results are obtained via the CLT, that is, when $L$ is larger, the analytical results are more accurate.
where the simulation results are used to verify the correctness of the analytical results.
In addition, we observe that as the number of reflective elements increases, the system performs better in terms of reliability.
This is because when $L$ is larger, the reflected signal energy reaching the UE is stronger and therefore the performance becomes better.

\begin{figure}[t]
\begin{minipage}[t]{0.49\linewidth}
\centering
\includegraphics[width=4.5 cm]{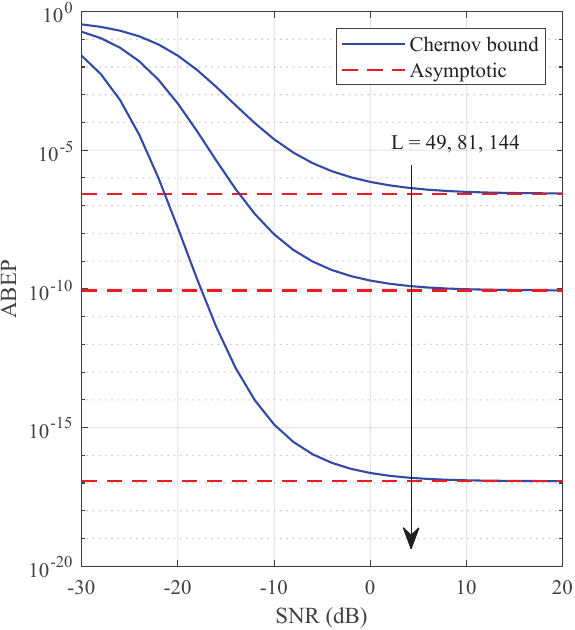}
\caption{\small{Validation of analytical derivation results under intelligent RIS-SSK scheme.}}
\label{intasy}
\end{minipage}%
\hfill
\begin{minipage}[t]{0.49\linewidth}
\centering
\includegraphics[width=4.5 cm]{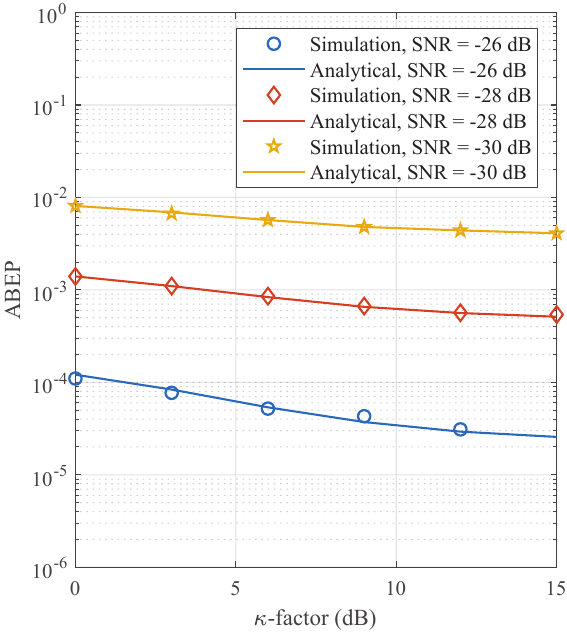}
\caption{\small{Impact of Rician factor $\kappa$ on ABEP under intelligent RIS-SSK scheme.}}
\label{kapver}
\end{minipage}
\end{figure}

Fig. \ref{verupbound} depicts the correctness of the Q-function approximating method on the ABEP expression and the Chernov bound on the ABEP expression.
Note that the Rician factor $\kappa$ and estimation error variance $\sigma_e^2$ are set to fixed values of 3 dB and 0.1, respectively.
As expected, the number of RIS reflection elements can effectively enhance the ABEP performance of the RIS-SSK scheme in the presence of imperfect CSI.
This is because the higher the number of RIS elements the stronger the signal power of the reflected convergence to the UE side.
In addition, the Q-function approximating results are quite close to those obtained from the exact integral form, which effectively verifies the tightness of closed-form expression.
In particular, the value of upper bound is significantly higher than the exact result because in (\ref{pep6}) we go to the maximum value $\pi/2$ at any point within $[0,\pi/2)$. After accumulating down the entire interval, the value of upper bound is overall higher than exact values in Fig. \ref{verupbound}.
On the other hand,
Fig. \ref{intasy} has the same parameter configuration as Fig. \ref{verupbound}. It is observed that there is not only a significant ABEP performance degradation but also an error floor of the RIS-SSK scheme with imperfect CSI.
This is because, by this time, the main factor that impacts reliability is not the AWGN anymore, but the noise comes from the channel estimation error.

In Fig. \ref{kapver}, we exhibit the performance impact of the LoS path of the reflection channel on the RIS-SSK scheme with imperfect CSI, where the number of reflecting elements is 144, and the estimation error variance of each reflecting element up to the UE is 0.1.
In particular, the analytical values are generated by the GCQ method. From Fig. \ref{kapver}, the simulation and analytical values match very well. The error can be reduced by increasing the number of simulations where there is no perfect overlap.
Obviously, when the Rician factor is larger, indicating a stronger signal energy for the reflected LoS path, the quality of the received signal on the UE side is higher. Hence, the ABEP performance can be improved.
Additionally, it is also found that the ABEP performance of RIS-SSK in the imperfect CSI case is enhanced as the SNR increases.

\begin{figure}[t]
\begin{minipage}[t]{0.49\linewidth}
\centering
\includegraphics[width=4.5 cm]{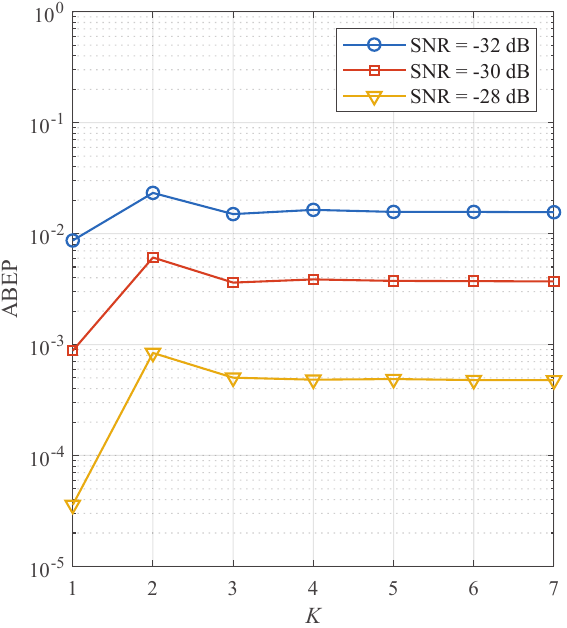}
\caption{\small{Convergence analysis of GCQ method.}}
\label{clti0}
\end{minipage}%
\hfill
\begin{minipage}[t]{0.49\linewidth}
\centering
\includegraphics[width=4.5 cm]{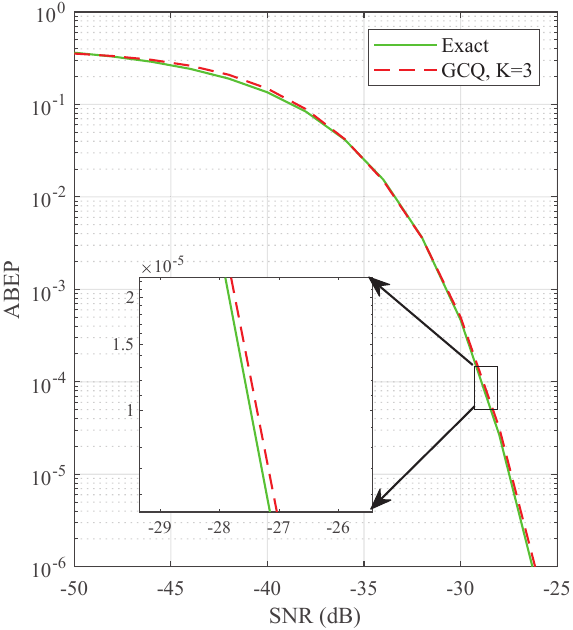}
\caption{\small{Accuracy analysis of GCQ method.}}
\label{clti}
\end{minipage}
\end{figure}

Fig. \ref{clti0} illustrates the variation of ABEP values with complexity-accuracy trade-off factor $K$ for SNR = -32 dB, -30 dB, and -28 dB, respectively, obtained using the GCQ method represented by (\ref{eqGCQ1}).
In particular, the parameters $\kappa$, $\sigma_e^2$, and $L$ are respectively defined as 3 dB, 0.1, and 200.
It is observed that the ABEP values obtained using the GCQ approach tend to converge in the case of $K=3$, and the ABEP values remain constant as the value of $K$ increases. On the other hand, to verify the gap between the ABEP obtained using the GCQ approach and the exact ABEP, we plotted Fig. \ref{clti}, where the remaining parameters remain the same as Fig. \ref{clti0} except for $K=3$.
As shown in Fig. \ref{clti}, when $K=3$, the difference between the result obtained by GCQ and the actual value is almost negligible. Consequently, the GCQ approach can achieve better system performance at a lower complexity $K=3$.

\begin{figure}[t]
\begin{minipage}[t]{0.49\linewidth}
\centering
\includegraphics[width=4.5 cm]{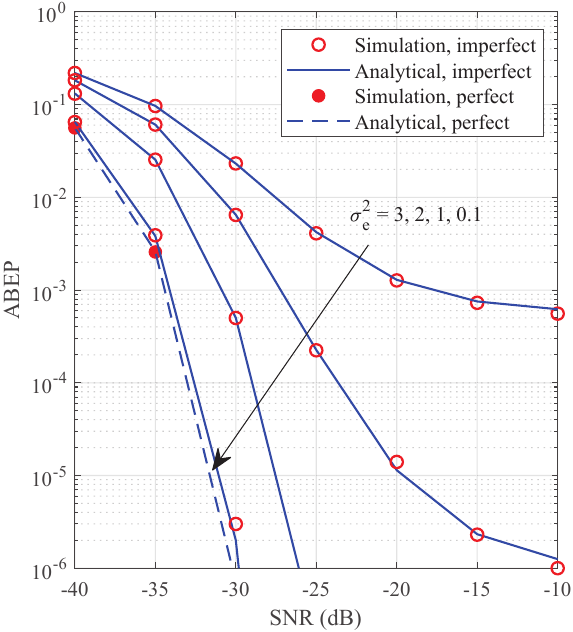}
\caption{\small{ABEP performance of intelligent RIS-SSK with fixed $\sigma_e^2$.}}
\label{fixsigma}
\end{minipage}%
\hfill
\begin{minipage}[t]{0.49\linewidth}
\centering
\includegraphics[width=4.5 cm]{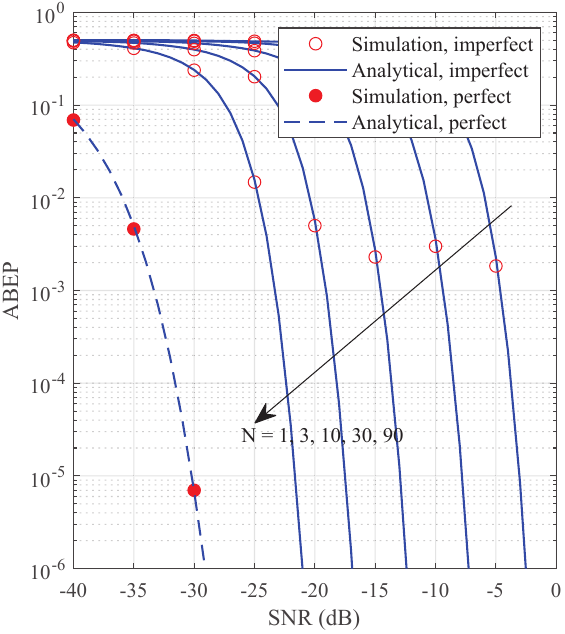}
\caption{\small{ABEP performance of intelligent RIS-SSK with variable $\sigma_e^2$.}}
\label{figvarpar}
\end{minipage}
\end{figure}

In Fig. \ref{fixsigma}, the Monte Carlo simulation results and analytical curves of the RIS-SSK scheme with $\kappa$ = 3 dB and $L=256$ are given, where the analytical curves are generated by the GCQ method.
Note that the error of channel estimation is set as fixed values $\sigma_e^2 = 3,2,1,0.1$, respectively.
That is, the correlation coefficients are $\zeta = 0.500,0.5774,0.7071,0.9535$.
As a reference, the corresponding ABEP with the perfect CSI is also shown with a dashed line for the RIS-SSK scheme.
First, as can be seen from Fig. \ref{fixsigma}, the analytical curves provided by (\ref{eqGCQ1}) become extremely tight with increasing SNR for all $\sigma_e^2$ values.
Second, we observed from Fig. \ref{fixsigma} that  a high correlation between the estimated channel and the real channel leads to a low error bit rate, which is also in line with our expectation.

In Fig. \ref{figvarpar}, simulation and analytical results of the intelligent RIS-SSK scheme are presented in the presence of imperfect CSI with respect to variable $\sigma_e^2$ values, where the number of pilots is chosen as $N$ = 1, 2, 10, 30, and 90.
Meanwhile, the number of reflective units of RIS and the Rician factor are set to $L = 256$ and $\kappa = 3$ dB, respectively.
It is worth noting that the intelligent RIS-SSK scheme under the perfect CSI is utilized as a reference scheme for comparative analysis.
As can be seen in Fig. \ref{figvarpar}, for this configuration, the simulation values and the analytical derivations of the RIS-SSK scheme again closely match.
In Fig. \ref{figvarpar}, the channel estimation error variance is considered to be inversely proportional to the SNR and the number of pilots, i.e., $1/({N\rho})$, instead of $\sigma_e^2$ being fixed independent of the SNR as considered in Fig. \ref{fixsigma}.
Accordingly, as the number of transmitted pilots increases, the channel is estimation improves, and the estimation error declines.

\begin{figure}[t]
\begin{minipage}[t]{0.49\linewidth}
\centering
\includegraphics[width=4.5 cm]{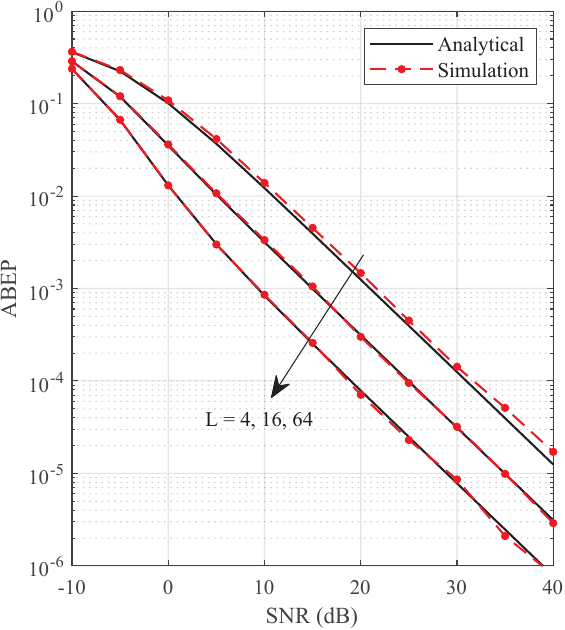}
\caption{\small{Verification of the conditions for the application of CLT under the blind RIS-SSK scheme.}}
\label{versablind}
\end{minipage}%
\hfill
\begin{minipage}[t]{0.49\linewidth}
\centering
\includegraphics[width=4.5 cm]{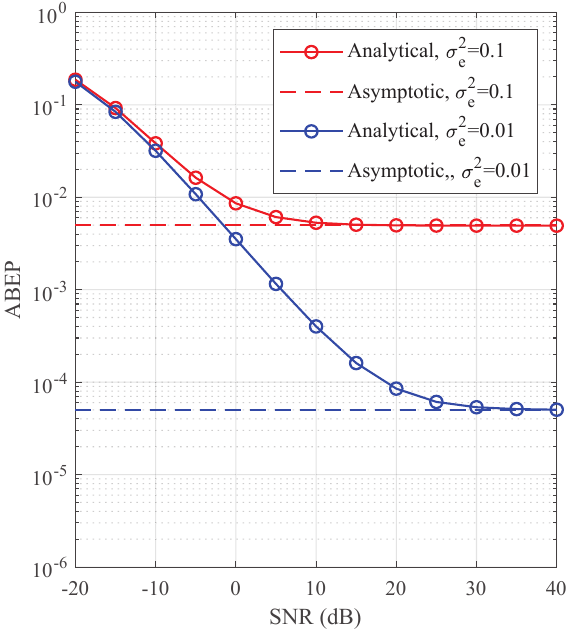}
\caption{\small{Verification of analytical derivation results under blind RIS-SSK scheme.}}
\label{blindasy}
\end{minipage}
\end{figure}

In Fig. \ref{versablind}, we plot the corresponding simulation results and analytical curves for the blind RIS-SSK system with variable channel estimation parameter $\sigma_e^2=1/(10\rho)$ and Rician factor $\kappa = 3$ dB.
Note that when SNR $>$ 30 dB, each simulation value is obtained by averaging $1\times 10^7$ channel generations.
As expected, the increasing number of RIS-equipped elements significantly enhances the ABEP performance of the blind RIS-SSK system.
Although RIS cannot adjust the phase shift of the transceiver signal, each reflector acts as an independent scatterer. As the number of reflective units increases, the signal strength reaching the UE can be effectively raised, thus improving system performance.
It can be observed from the Fig. \ref{versablind} that, in this scheme, the analytical results obtained by CLT almost coincide perfectly with the simulation results for the number of RIS elements of 16.

Fig. \ref{blindasy} shows the ABEP of the blind RIS-SSK system in the absence of CSI as a function of the average SNR (i.e., $\rho$) under the $L = 144$ scenario, where the Rician factor between each reflecting element and the UE-side are set as $\sigma_e^2=0.1$ and $\sigma_e^2=0.01$, respectively.
Note that the asymptotic ABEP results are generated via the (\ref{asybli}).
From Fig. \ref{blindasy}, it can be seen that the analytical expression of ABEP with $L=144$ and its corresponding asymptotic expression are in excellent agreement in the high SNR region, which is a powerful confirmation of the correctness of the derived asymptotic ABEP expression.
However, in the low SNR region, the gap between the two is extremely large, due to the fact that the asymptotic ABEP derived in (\ref{asybli}) represents the trend of ABEP at high SNR.

\begin{figure}[t]
\centering
\includegraphics[width=6cm]{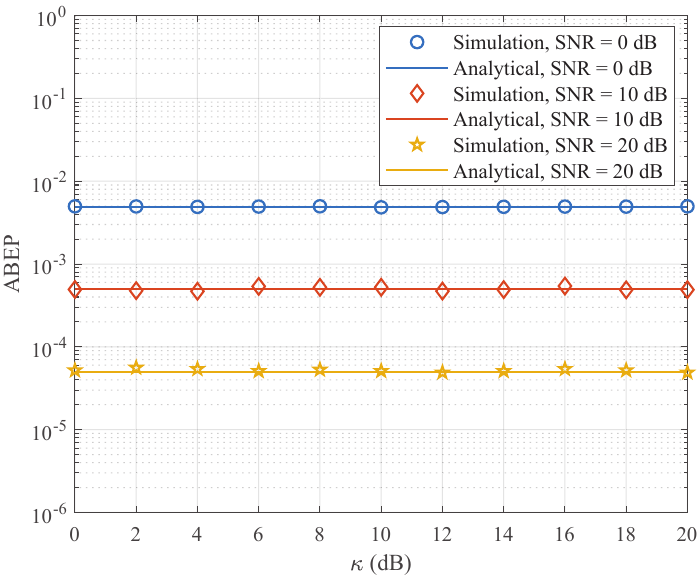}
\caption{\small{Impact of Rician factor $\kappa$ on ABEP under blind RIS-SSK scheme.}}
\vspace{-10pt}
\label{kappablind}
\end{figure}
In Fig. \ref{kappablind}, we depict the performance impact of the RIS and UE-side channel Rician factors on the blind RIS-SSK system, where the simulation results and the analysis curve of ABEP are displayed when $\kappa$ goes from 0 dB to 20 dB for three cases of SNR = 0 dB, 10 dB, and 20 dB, respectively.
Note that the number of reflection elements of RIS is set to $L=100$.
The simulation results and the analytical curves match each other, which further verifies the correctness of the analytical derivation. In addition, there are very few points that do not perfectly match because the number of simulations is not enough, and this phenomenon can be effectively improved by increasing the number of simulations.
Moreover,
as $\kappa$ increases, the value of ABEP remains constant. From (\ref{ricdem}), we can get that the Rician component is canceled out. In other words, in the blind RIS-SSK system, the Rician fading channel degenerates to a Rayleigh fading channel.

\begin{figure}[t]
\begin{minipage}[t]{0.49\linewidth}
\centering
\includegraphics[width=4.5 cm]{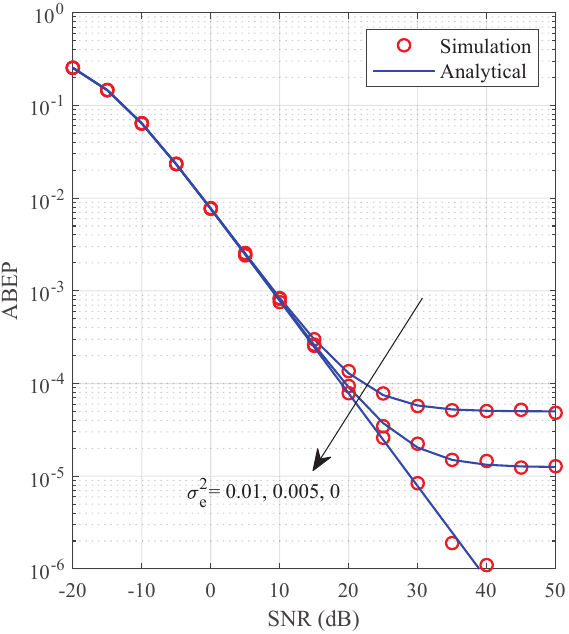}
\caption{\small{ABEP performance of blind RIS-SSK with fixed $\sigma_e^2$.}}
\label{blind_ana1}
\end{minipage}%
\hfill
\begin{minipage}[t]{0.49\linewidth}
\centering
\includegraphics[width=4.5 cm]{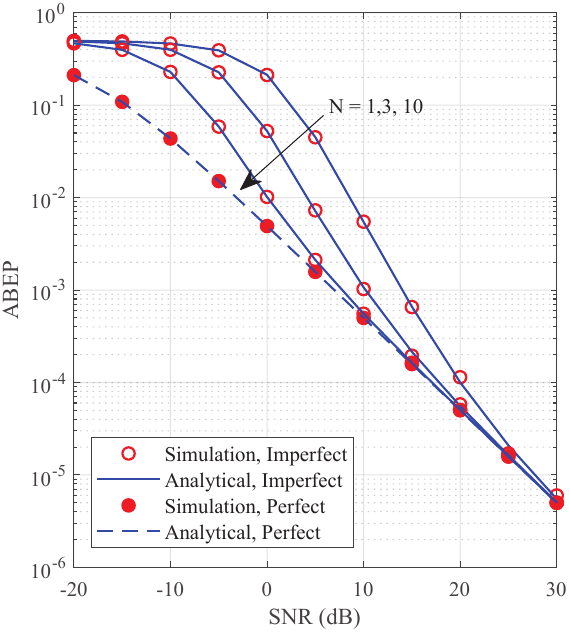}
\caption{\small{ABEP performance of blind RIS-SSK with variable $\sigma_e^2$.}}
\label{blind_ana2}
\end{minipage}
\end{figure}

Fig. \ref{blind_ana1} depicts the ABEP performance of the blind RIS-SSK scheme in the presence of channel estimation errors, where the number of the reflecting elements and Rician factor are configured as $L=100$ and $\kappa=5$ dB, respectively.
It is observed that the analytical upper bounds of ABEP are tight in all SNR regions.
This is because the result of the analytical upper bound of ABEP in the case of  $N_t=2$ is equivalent to the true value of ABEP.
Meanwhile, it is seen that the blind RIS-SSK scheme under the perfect CSI achieves better ABEP performance than that of imperfect CSI.
To be specific, at an SNR value of 30 dB, when compared with perfect CSI ABEP= $0.8 \times 10^{-6}$ case $(\sigma_e^2=0)$, the ABEP of the blind RIS-SSK scheme is $0.1\times 10^{-5}$ and $0.5\times 10^{-5}$ under the $\sigma_e^2=0.005$ and $\sigma_e^2=0.01$, respectively.

In Fig. \ref{blind_ana2}, we illustrate the impact of channel estimation error on the ABEP performance of blind RIS-SSK scheme with variable $\sigma_e^2$, that is, the number of transmitted pilots is $N$ = 1, 3, and 10.
It is evident that we validate the accuracy of the PEP of the ML detector and the tightness of the ABEP upper bound in Fig. \ref{blind_ana2}.
As expected, the ABEP performance of the blind RIS-SSK scheme in the presence of imperfect CSI improves as the number of transmit pilots increases.
In addition, we find an interesting phenomenon that the system performance of imperfect CSI gradually converges to perfect CSI as the SNR increases. This is because the variable $\sigma_e^2$ is not only related to the number of leads but also to the SNR. Accordingly, as the SNR is much larger than $N$, the effect of $N$ is almost negligible.

\begin{figure}[t]
\begin{minipage}[t]{0.49\linewidth}
\centering
\includegraphics[width=4.5 cm]{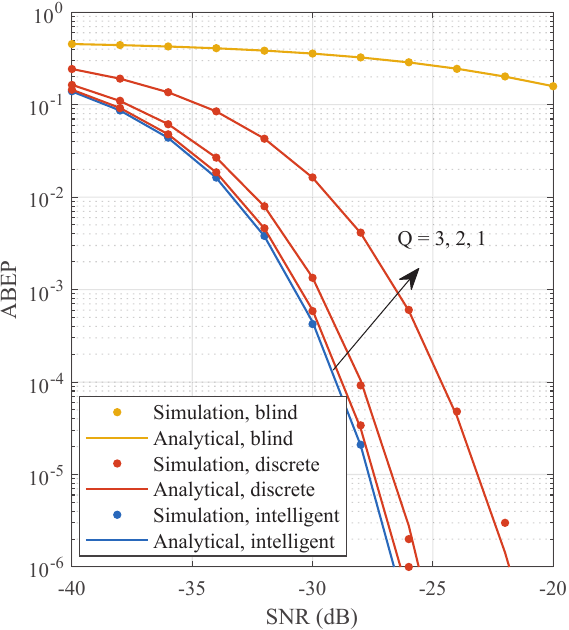}
\caption{\small{Impact of bit quantization on RIS-SSK under fixed $\sigma_e^2$.}}
\label{bitfix}
\end{minipage}%
\hfill
\begin{minipage}[t]{0.49\linewidth}
\centering
\includegraphics[width=4.5 cm]{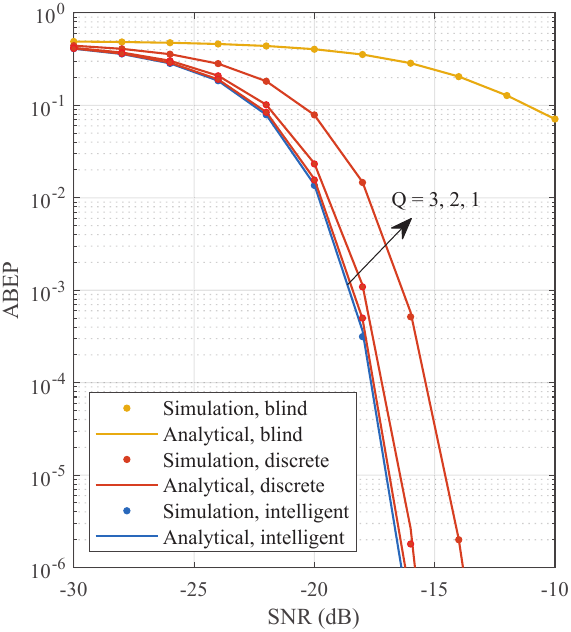}
\caption{\small{Impact of bit quantization on RIS-SSK under variable $\sigma_e^2$.}}
\label{bitvar}
\end{minipage}
\end{figure}

In Fig. \ref{bitfix}, we consider a more practical operating mode of the RIS-SSK scheme in the case of perfect CSI, i.e., RIS with 1, 2, and 3 bits discrete phase shifts, respectively.
To clarify the impact of discrete phase shifts on the system performance, we also provide two reference schemes for RIS with intelligent transmit phase shift and blind reflection phase shift, where the variance of the estimation error adopts a fixed value $\sigma_e^2=0.1$.
Note that the parameters with respect to the Rician factor and the number of reflecting elements are set as $\kappa = 3$ dB and $L = 196$, respectively.
A substantial performance improvement is obtained when 1-bit quantization is employed compared to the blind scheme. Additionally, it is found that the ABEP performance curve associated with  3-bit quantization is very close to the ABEP value with the continuous phase.

In Fig. \ref{bitvar}, we investigate the performance of RIS at 1 bit, 2 bit, and 3 bit quantization phase shifts on the RIS-SSK scheme at imperfect CSI, respectively. It is worth mentioning that in the previous figure, we can observe that the error curves achieved by 3 bit quantization and continuous phase shifting are quite close to each other. Hence, we also adopt the $1\times 10^7$ channel generations to verify the analytical results by Monte Carlo simulation.
Besides, the number of pilots, Rician factor, and the number of reflecting elements are set as $N=30$, $\kappa=3$ dB, and $L=196$, respectively.
It can be observed from Fig. \ref{bitvar} that the blind RIS-SSK scheme has the worst performance.
On the contrary, the intelligent RIS-SSK scheme has the best performance. In addition, the performance of the discrete phase improves as the number of quantized bits increases.
It is worth mentioning that the ABEP values obtained by 3-bit quantization are remarkably close to the ideal reflection phase shift of the RIS.
\section{Conclusion}
This paper presented the ABEP performance of RIS-SSK with imperfect channel estimation. The channel between the BS-RIS and RIS-UE is subject to Rayleigh fading, while the channel between the RIS-UE is subject to Rician fading. Two schemes, namely the intelligent RIS-SSK scheme and the blind RIS-SSK scheme, are considered with channel estimation errors. We derive the PDF of the non-central chi-square distribution with one degree of freedom and the exact integral ABEP of the intelligent RIS-SSK scheme based on the ML detector. Additionally, we derive the ABEP expression vi the GCQ approach and obtain closed-form expressions relying on Q-function approximating approach. Also, the asymptotic ABEP expression is provided. For the blind RIS-SSK scheme, we derive closed-form analytical ABEP  expression and asymptotic ABEP expression under imperfect CSI. In particular, the discrete phase is also investigated. Through simulations, it demonstrates that the precision of the analytical derivation of the ABEP expression and the accuracy of the average ABEP in both schemes.  It is shown that the intelligent and blind RIS-SSK schemes represent the two limiting forms of the RIS quantization scheme. It is worth mentioning that the composite channel is unaffected by the Rician factor in the blind RIS-SSK scheme.
For future work, path loss can be considered, so that the choice of location for RIS deployment becomes an optimization problem. Further, RIS aided dual-polarized shifting keying scheme is also an open and interesting research problem.

\begin{appendices}

\section{Proof of Lemma 1}
Without loss of generality, we let $z = \eta-\hat\eta$. Recalling Eq. (\ref{musigma}), it is known that $z$ follows the true Gaussian distribution. Thus, the PDF of $z$ can be calculated as
\begin{equation}
f_Z(z)=\frac{1}{\sqrt{2\pi\sigma^2}}\exp\left({-\frac{\left(z-\mu\right)^2}{2\sigma^2}}\right).
\end{equation}
Since the variable $X$ satisfies $X=Z^2$, the CDF of the parameter $X$ can be evaluated as
\begin{equation}\label{cdfx01}
\begin{aligned}
F_X(x) &= \Pr(X\leq x)= \Pr(Z^2\leq x)= \Pr(-\sqrt{x}\leq Z\leq \sqrt{x})\\
&=\frac{1}{\sqrt{2\pi\sigma^2}}\int_{-\sqrt{x}}^{\sqrt{x}} \exp\left({-\frac{(z-\mu)^2}{2\sigma^2}}\right)dz.
\end{aligned}
\end{equation}
Based on Eq. (\ref{cdfx01}), the PDF of $X$ can be expressed as
\begin{equation}\label{pdfX}
\begin{aligned}
f_X(x)&=\frac{d F_X(x)}{dx}=\frac{1}{2\sqrt{2x\pi\sigma^2}}\\&\times\left[\exp\left({-\frac{(\sqrt{x}-\mu)^2}{2\sigma^2}}\right)+\exp\left({-\frac{(-\sqrt{x}-\mu)^2}{2\sigma^2}}\right)\right].
\end{aligned}
\end{equation}
After some manipulations, we have
\begin{equation}
\begin{aligned}
f_X(x)
&=\frac{\exp\left({-\frac{x+\mu^2}{2\sigma^2}}\right)}{2\sqrt{x}\sqrt{2\pi\sigma^2}}\left[\exp\left({\frac{\sqrt{x}\mu}{\sigma^2}}\right)+\exp\left({-\frac{\sqrt{x}\mu}{\sigma^2}}\right)\right].
\end{aligned}
\end{equation}
Herein, the proof of {\bf Lemma 1} is completed.

\section{}
For the $\Phi (x)$ denotes Gaussian error function, the mathematical expression can be given by
\begin{equation}\label{eqphi2}
    \Phi(x)=\frac{2}{\sqrt{\pi}}\int_0^x\exp(-v^2)dv.
\end{equation}
Then, let us set
$
\Phi(-x)=\frac{2}{\sqrt{\pi}}\int_0^{-x}\exp(-v^2)dv.
$
Let us define $z=-v$, we have
\begin{equation}\label{eqphi1}
\Phi(-x)=\frac{2}{\sqrt{\pi}}\int_0^{-x}\exp(-z^2)dz.
\end{equation}
Combining Eqs. (\ref{eqphi2}) and (\ref{eqphi1}), we have
$
\Phi(-x) + \Phi(x) = 0.
$
Replace the corresponding parameter $\sqrt{{\frac{{\mu^2({1+\rho(1-\zeta^2)\sigma_e^2L})\sigma^2\sin^2\o}}{{2\sigma^4({1+\rho(1-\zeta^2)\sigma_e^2L})\sin^2\o+\rho\zeta^2\sigma^6}}}}$ with $x$, the proof of {\bf Theorem 1} is completed.

\section{Proof of Lemma 2}
By observing Eq. (\ref{qapr1}), we find that it consists of the sum of two similar terms.
For brevity, we set the coefficients before the variable $x$ to $s$.
Thus, the MGF function of $x$ can be described as
\begin{equation}\label{mgf1}
\begin{aligned}
M_X(s)&=\int_0^\infty \exp(sx)f(x)dx.
\end{aligned}
\end{equation}
Substituting Eq. (\ref{pdfX}) into Eq. (\ref{mgf1}), $M_X(s)$ can be stated as
\begin{equation}\label{mgf2}
\begin{aligned}
M_X(s)
&=\int_0^\infty \frac{\exp(sx)\exp\left(-\frac{x+\mu^2}{2\sigma^2}\right)}{2\sqrt{2\pi\sigma^2 x}}\left[\exp\left(\sqrt{\frac{\mu^2x}{\sigma^4}}\right)\right.\\&\left.+\exp\left(-\sqrt{\frac{\mu^2x}{\sigma^4}}\right)\right]dx.
\end{aligned}
\end{equation}
After some substitution, Eq. (\ref{mgf2}) can be simplified to
\begin{equation}
\begin{aligned}
M_X(s)
&=\frac{\exp\left(-\frac{\mu^2}{2\sigma^2}\right)}{\sqrt{2\pi\sigma^2 }}\int_0^\infty \exp\left(-\frac{1-2\sigma^2s}{2\sigma^2}x^2\right) \\& \times\left[\exp\left({\frac{\mu x}{\sigma^2}}\right)+\exp\left(-\frac{\mu x}{\sigma^2}\right)\right]dx.
\end{aligned}
\end{equation}
With some simple mathematical operations, $M_X(s)$ can be evaluated as
\begin{equation}\label{mgf4}
\begin{aligned}
M_X(s)
&=\sqrt{\frac{1}{1-2s\sigma^2}}\exp\left(\frac{s\mu^2}{1-2s\sigma^2}\right).
\end{aligned}
\end{equation}
Upon substituting (\ref{mgf4}) into (\ref{qapr1}), the proof of {\bf Lemma 2} is completed.

\end{appendices}

\end{document}